\documentclass[twocolumn,secnumarabic,amssymb, nobibnotes, 
aps, prl,superscriptaddress, showpacs, showkeys, floatfix]{revtex4-1}
\usepackage{graphicx}
\usepackage{array}
\usepackage{csvsimple,longtable,booktabs}
\usepackage{multirow}
\usepackage{adjustbox}
\usepackage{hyperref,isotope,amsmath,mhchem}

\begin{document}
% 
% \preprint{APS/123-QED}
% 
\title{\boldmath 
\isotope[138]Ba$(d,\alpha)$ study of states in $^{136}{\rm Cs}$: Implications for new physics searches with xenon detectors}% Force line breaks with \\
\author{B.\,M.~Rebeiro}
%\email{b.rebeiro@gmail.com}
\affiliation{Department of Physics and Astronomy, University of the Western Cape, P/B X17, Bellville 7535, South Africa}%
\affiliation{Department of Physics, McGill University,
Montr\'eal, Qu\'ebec H3A 2T8, Canada}
\author{S.~Triambak}
\email{striambak@uwc.ac.za}
\affiliation{Department of Physics and Astronomy, University of the Western Cape, P/B X17, Bellville 7535, South Africa}%
\author{P.\,E.~Garrett}
\affiliation{Department of Physics, University of Guelph, Guelph, Ontario N1G 2W1, Canada}%
\affiliation{Department of Physics and Astronomy, University of the Western Cape, P/B X17, Bellville 7535, South Africa}%
\author{G.\,C.~Ball}
\affiliation{TRIUMF, 4004 Wesbrook Mall, Vancouver, British Columbia V6T 2A3, Canada}%
 \author{B.\,A.~Brown}
\affiliation{Department of Physics and Astronomy and National Superconducting Cyclotron Laboratory, Michigan State University, East Lansing, Michigan 48824-1321, USA}%
\author{J.~Men\'endez}
\affiliation{Department of Quantum Physics and Astrophysics and Institute of Cosmos Sciences,
University of Barcelona, 08028 Barcelona, Spain}
\author{B.~Romeo}
\affiliation{Donostia International Physics Center, 20018 San Sebasti\'an, Spain}
\author{P.~Adsley}
\affiliation{Cyclotron Institute and Department of Physics \& Astronomy, Texas A\&M University, College Station, Texas 77843, USA}%
\author{B.\,G.~Lenardo}
\affiliation{SLAC National Accelerator Laboratory, Menlo Park, California 94025, USA}%
\author{R.~Lindsay}
\affiliation{Department of Physics and Astronomy, University of the Western Cape, P/B X17, Bellville 7535, South Africa}%
\author{V.~Bildstein}
\affiliation{Department of Physics, University of Guelph, Guelph, Ontario N1G 2W1, Canada}%
\author{C.~Burbadge}
\altaffiliation[Deceased.]{}
\affiliation{Department of Physics, University of Guelph, Guelph, Ontario N1G 2W1, Canada}%
\author{R.~Coleman}
\affiliation{Department of Physics, University of Guelph, Guelph, Ontario N1G 2W1, Canada}%
\author{A.~Diaz Varela}
\affiliation{Department of Physics, University of Guelph, Guelph, Ontario N1G 2W1, Canada}%
\author{R.~Dubey}
\affiliation{Department of Physics and Astronomy, University of the Western Cape, P/B X17, Bellville 7535, South Africa}%
\author{T.~Faestermann }
\affiliation{Physik Department, Technische Universit\"{a}t M\"{u}nchen, D-85748 Garching, Germany}%
\author{R.~Hertenberger}
\affiliation{Fakult\"{a}t f\"{u}r Physik, Ludwig-Maximilians-Universit\"{a}t M\"{u}nchen, D-85748 Garching, Germany}%
\author{M.~Kamil}
\affiliation{Department of Physics and Astronomy, University of the Western Cape, P/B X17, Bellville 7535, South Africa}%
 \author{K.\,G.~Leach}
\affiliation{Department of Physics, Colorado School of Mines, Golden, Colorado 80401, USA}
\author{C.~Natzke}
\affiliation{Department of Physics, Colorado School of Mines, Golden, Colorado 80401, USA}
\author{J.\,C.~Nzobadila Ondze}
\affiliation{Department of Physics and Astronomy, University of the Western Cape, P/B X17, Bellville 7535, South Africa}
\author{A.~Radich}
\affiliation{Department of Physics, University of Guelph, Guelph, Ontario N1G 2W1, Canada}%
\author{E.~Rand}
\affiliation{Department of Physics, University of Guelph, Guelph, Ontario N1G 2W1, Canada}%
\author{H.\,-F.~Wirth}
\affiliation{Fakult\"{a}t f\"{u}r Physik, Ludwig-Maximilians-Universit\"{a}t M\"{u}nchen, D-85748 Garching, Germany}%
\date{\today}% It is always \today, today,

 \begin{abstract}
We used the \isotope[138]Ba$(d,\alpha)$ reaction to carry out an in-depth study of states in \isotope[136]{Cs}, up to around 2.5~MeV. In this work, we place emphasis on hitherto unobserved states below the first $1^+$ level, which are important in the context of solar neutrino and fermionic dark matter (FDM) detection in large-scale xenon experiments. We identify for the first time candidate metastable states in \isotope[136]{Cs}, which would allow a real-time detection of solar neutrino and FDM events in xenon detectors, with high background suppression. Our results are also compared with shell-model calculations performed with three Hamiltonians that were previously used to evaluate the nuclear matrix element (NME) for \isotope[136]Xe neutrinoless double beta decay. We find that one of these Hamiltonians, which also systematically underestimates the NME compared to the others, dramatically fails to describe the observed low-energy \isotope[136]Cs spectrum, while the other two show reasonably good agreement.
 \end{abstract}
\maketitle
% 
% \tableofcontents
% 
It has been pointed out~\cite{Raghavan} that double beta decaying atomic nuclei provide the necessary framework to perform real-time spectroscopic studies of solar neutrinos, with high background suppression.
%In such cases the pairing interaction renders the even-even parent nucleus (with spin-parity $J^\pi = 0^+$) more bound than its $(A,Z+1)$ neighbor, and precludes single $\beta$ transitions of the type $(A,Z) \to (A,Z+1)$. This scenario  presents a `stable' target for the solar $\nu_e$ flux, $\phi_e$, and also results in low thresholds for charged-current (CC) $\nu_e$ capture to low-lying $1^+$ states in the intermediate $(A,Z+1)$ nucleus. As the intermediate nucleus is odd-odd, its low-lying structure is mainly defined by two-quasiparticle configurations for the unpaired proton and neutron. Such configurations typically lead to the existence of metastable states, whose long half-lives permit a nearly background-free identification of CC solar $\nu_e$ captures, via a delayed coincidence analysis~\cite{Raghavan}.
%differnt 2 qp configurations, K selection rules, smallness of the matrix elements, delta E... so many qp configs leads to many levels.
%javier
In such cases, the parent nucleus has an even number of protons $(Z)$ and neutrons $(N)$, with $A = Z + N$, and total angular momentum-parity $J^\pi=0^+$. Consequently, the attractive nuclear pairing interaction renders it more bound than its isobaric $(A,Z+1)$ neighbor, which has odd $Z$ and $N$. This scenario precludes single $\beta$ transitions of the type $(A,Z) \to (A,Z+1)$ and
presents a `stable’ target for the solar $\nu_e$ flux, $\phi_e$. It also results in low thresholds for charged-current (CC) $\nu_e$ capture to $J^\pi = 1^+$ states in the $(A, Z+1)$ system.  As this intermediate nucleus is odd-odd, its low-lying structure is mainly defined by two-quasiparticle configurations for the unpaired proton and neutron. Such configurations may lead to the existence of metastable states, with long half-lives that permit a nearly background-free identification of CC solar $\nu_e$ captures, via a delayed coincidence analysis~\cite{Raghavan}.

In this regard, xenon-based detectors~\cite{Adhikari:21,kamland,next,pandax:22,Akerib:21, Macolino,Aprile:22} present an opportunity for solar neutrino detection, both at the tonne-scale and beyond. As examples, the nEXO~\cite{Adhikari:21}, KamLAND-Zen~\cite{kamland} and NEXT~\cite{next} experiments rely on isotopically-enriched xenon to search for lepton-number-violating (LNV) neutrinoless double beta decays ($0\nu2\beta$) of \isotope[136]{Xe}. The low  $\nu_e$ reaction threshold for \isotope[136]{Xe} presents a compelling case to use such xenon detectors for solar neutrino astronomy at energies $\lesssim 1$~MeV.  A previous study~\cite{Scott:20} showed that the dominant CC $\nu_e$ captures on \isotope[136]{Xe} will be through the two lowest-energy $1^+$ states in \isotope[136]Cs, at 591 and 845~keV respectively~\cite{Frekers2013},  with the former $1^+_1$ state being the most significant ($Q_\nu = 681.3$~keV). Therefore, detectors loaded with \isotope[136]{Xe} will be sensitive to $\phi_e (\mathrm{CNO},~\isotope[7]{Be},~ \isotope[8]{B},~pep)$. Of particular interest are \isotope[7]{Be} electron-capture neutrinos and those emitted from the solar CNO cycle, whose detection will offer insight into the innermost core of the Sun~\cite{Bachall:94,borexino1,borexino2}. Additionally, such experiments can also identify similar CC-type excitations to $1^+$ states in \isotope[136]{Cs}, caused via MeV-scale fermionic dark matter (FDM) absorption~\cite{Dror1,Dror2} on \isotope[136]{Xe}. %Such interactions are possible in a modified left-right symmetric model~\cite{Dror2}, where the electron and the FDM candidate constitute a right-handed doublet that transforms under $\mathrm{SU}(2)_R$.

Based on the above, a search for FDM absorption on \isotope[136]{Xe} was recently performed~\cite{Soud}. However, the analysis was %recently performed using the EXO-200 dataset~\cite{Soud}. However, the analysis was
severely challenged by the meager experimental information~\cite{Mccutchan2018} available for the low-lying level scheme of \isotope[136]Cs. Only three states
have thus far been experimentally verified below the $1_1^+$ level, with assigned $J^\pi$ values of $4^+$, $8^-$ and $9^-$, respectively~\cite{Mccutchan2018,Wimmer:11,Astier:13}. Independently, shell-model calculations~\cite{Scott:20} were performed to predict $\gamma$-ray deexcitation paths from the $1_1^+$ level in \isotope[136]Cs.
% which is expected to be predominantly populated via solar $\nu_e$ CC scattering.
The results showed promise for solar $\nu_e$ detection in both current and next-generation xenon experiments, mainly because of feeding to the predicted first excited state in \isotope[136]{Cs}  $(E_\mathrm{x} = 23~\mathrm{keV}; J^\pi = 3^+)$, which connects to the ground state via a slow ($\tau = 851$~ns~\footnote{We believe that the authors of Ref.~\cite{Scott:20} did not take into account internal conversion. They quote the partial $\gamma$-ray transition lifetime, $\tau_\gamma = 624~\mu$s, as the total lifetime of the level.}) $3_1^+ \to 5_1^+$ electric-quadrupole $(E2)$ transition. %The theory-predicted lifetime of this state was $\tau = 851$~ns~\footnote{We believe that the authors of Ref.~\cite{Scott:20} did not take into account internal conversion. They quote the partial $\gamma$-ray transition lifetime, $\tau_\gamma = 624~\mu$s, as the total lifetime of the level.}.
However, this level has not been experimentally validated to date. A more comprehensive elucidation of the low-lying structure of \isotope[136]{Cs} is essential to make further progress in this regard.

There is additional widespread interest to accurately determine the nuclear matrix elements (NMEs) for various $0\nu2\beta$ candidates, including \isotope[136]Xe~\cite{Engel_2017,Ejiri:19,bmf,Agostini}.  The calculated NME for this particular case ranges from $M^{0\nu} =1.11$--$4.77$~\cite{Rebeiro:20}, for light Majorana neutrino exchange. This theoretical limitation translates into an inevitable uncertainty band~\cite{kamland} on the LNV parameter responsible for the decay, which is hoped to be extracted from future experiments. Within the nuclear shell-model, the NME is in the range $M^{0\nu} = 1.63$--$2.45$, depending on the Hamiltonian used for the calculation. This spread is primarily because one of the Hamiltonians (QX) yields a systematically lower value for $M^{0\nu}$, by about 40\%, as shown in Table~\ref{tab:NMEs}. This systematic discrepancy persists~\cite{Lotta:23} even when recently acknowledged short-range NMEs~\cite{Cirigliano:18} are taken into consideration.
Therefore, an accurate understanding of the low-energy level scheme in \isotope[136]Cs also presents a robust testing ground for theory calculations of the \isotope[136]Xe $0\nu2\beta$ NME. This is because comparisons with experiment are much more sensitive to details of the nuclear Hamiltonian in odd-odd nuclei. Such details can be masked in even-even systems such as \isotope[136]Xe and \isotope[136]Ba, because of the dominant pairing interaction and other collective effects.
\begin{table}[t]
  \begin{flushleft}
% \begin{adjustbox}{width=1.0\textwidth}
  \caption{\label{tab:NMEs}
 Shell-model-evaluated NMEs for \isotope[136]Xe $0 \nu 2 \beta$.
 %Complete results for all states observed in this work (up to 2.5~MeV) are presented in the supplemental material.
  }
  \begin{ruledtabular}
  \begin{tabular}{cc}
  \multicolumn{1}{c }{Hamiltonian} &  \multicolumn{1}{c}{$M^{0\nu}$}  \\
  \colrule
  GCN5082~\cite{gcn} & 2.28, 2.45~\cite{javier}\\
  $V_{\mathrm{low}\text{-}k}$~\cite{coraggio} & 2.39~\cite{coraggio}\\
  JJ55t (SN100t)~\cite{jj55} & 2.06, 2.21~\cite{jj55}\\
  QX (SVD, MC)~\cite{qx} & 1.63, 1.76~\cite{Neascu:15}\\
  %\hline
  %\hline
  \end{tabular}
%$^a$~Possible unresolved doublet.
\end{ruledtabular}
\end{flushleft}
 \end{table}
  \begin{figure}[t]
   \includegraphics[width=0.49\textwidth]{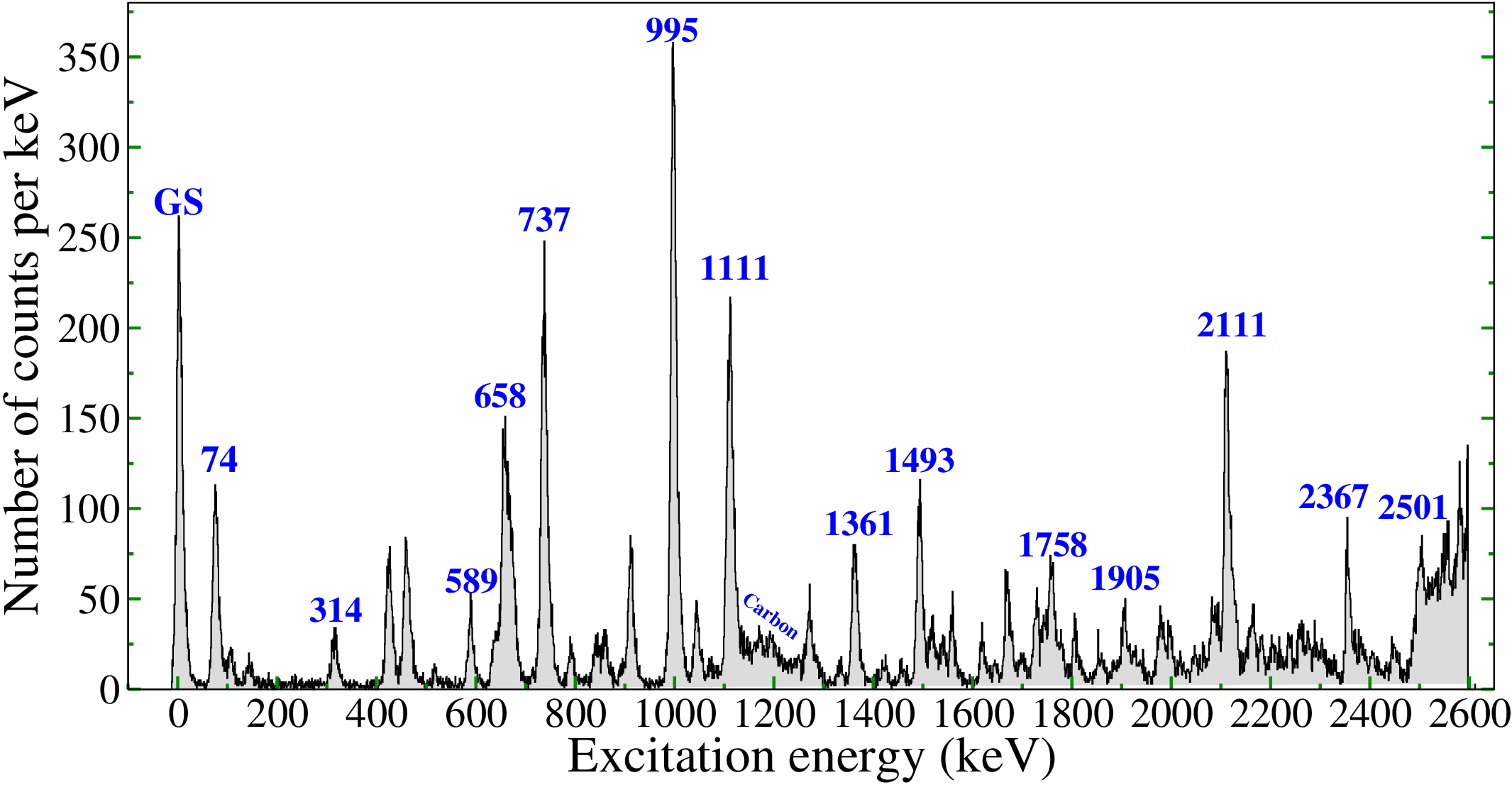}
   \caption{{\label{Fig:136Cs_calib}} Sample \isotope[138]Ba$(d,\alpha)$ spectrum obtained at $\theta_\mathrm{lab} = 10^\circ$. A few prominent peaks are labeled. %Previously identified states~\cite{Wimmer:11, Frekers2013, Puppe:11, Astier:13,Mccutchan2018} are labeled.
   }
  \end{figure}

With these motivations in place, this work reports a detailed high-resolution investigation of low-lying states in \isotope[136]Cs. %placing emphasis on multipoles with $J \ge 1$.
We used the \isotope[138]Ba$(d,\alpha)$\isotope[136]Cs two-nucleon transfer reaction, which is well suited for such a study.
%because of its selection rules and the large momentum mismatch between the incoming and outgoing particles that favors transitions to high $J$ states in \isotope[136]Cs.

  %\section{Experimental Details}\label{Sect:Expt_details}
The experiment was performed at the Maier-Leibnitz Laboratorium (MLL) in Garching, Germany. A 600~nA, 22~MeV deuteron beam was incident on a $99.8\%$ enriched $40~\mu$g/cm$^2$-thick $^{138}\rm{BaO}$ target, evaporated on a carbon foil.
  %Particles from the $^{138}$Ba$(d,\alpha)$ reaction 
The reaction ejectiles were momentum analyzed with the high-resolution Q3D magnetic spectrograph~\cite{LOFFLER19731}. % and focused onto its focal-plane detector, whose position resolution was $\approx$ 0.1 mm~\cite{HERTENBERGER1987201}.
The $\alpha$ particles were selected by comparing the partial energy losses of the reaction products in two gas proportional counters and the total energy deposited in a plastic scintillator detector at the focal plane.
  For energy calibration, we used the $^{94}$Mo$(d,\alpha)^{92}$Nb and $^{92}$Zr$(d,\alpha)^{90}$Y reactions on enriched $^{94}$MoO$_3$ and $^{92}$Zr targets that had thicknesses of $100~\mu$g/cm$^2$ and $50~\mu$g/cm$^2$,  respectively. %These energy calibrations were performed at $\theta_{lab}=10^\circ$, and
  The calibrations
  %used well-known states in $^{90}$Y~\cite{90Y} and $^{92}$Nb~\cite{92Nb}, and 
  explicitly took into account differences in reaction kinematics and energy losses within the target foils, as described in~Refs.~\cite{Mukwevho2018, BernadetteThesis}.
  A sample calibrated \isotope[138]Ba$(d,\alpha)$ spectrum is shown in Fig.~\ref{Fig:136Cs_calib}. The measured full widths at half maxima (FWHM) of the $\alpha$ peaks were $\sim 10$~keV, vastly superior than the 40~keV resolution reported in a previous \isotope[136]Xe$(\isotope[3]He,t)$ study~\cite{Puppe:11,Frekers2013} that mainly investigated $1^+$ states in \isotope[136]Cs.
  
  The \isotope[138]Ba$(d,\alpha)$ spectra were collected at different angles in the range $\theta_{lab}=5^\circ-45^\circ$, at $5^\circ$ intervals. Additionally, \isotope[138]Ba$(d,d)$ elastic scattering data were acquired in the range $\theta_{lab}=15^\circ -115^\circ$, at $5^\circ$ intervals. We used these datasets to determine the target thickness and obtain differential scattering cross sections, as described in Refs.~\cite{Rebeiro:20,Rebeiro:21}. The measured angular distributions were then compared to distorted wave Born approximation (DWBA) predictions, provided by the DWUCK5 computer code.

%Before delving into details of the data analysis, it seems appropriate at this point to highlight some salient features of the $(d,\alpha)$ deuteron-transfer reaction on an even-even target, with $J^\pi = 0^+$. 

The selectivity of the $(d,\alpha)$ reaction is such that the transferred $np$ pair is in a relative $l = 0$ state, with spin $S = 1$ and isospin $T = 0$~\cite{Glendenning:63}.
%Previous work has shown that both nucleons are preferentially picked up from the same single-particle ($j^2$) configuration, with their angular momenta coupled to the maximum value $J = 2j$~\cite{Harvey:62,Rivet:66,Lu:69}.
If both nucleons are picked up from the same single-particle ($j^2$) configuration,
%In such a scenario,
%the orbital angular momentum $L$ transferred in the reaction is even. In such a scenario,
the total angular momentum $J$ of the final state is necessarily odd. However, if
the neutron and proton are picked up from different configurations, with $\boldsymbol{L = l_n + l_p}$, then
$J = L$ and $J = L \pm 1$ states, with parity $(-1)^{l_n+l_p}$ are produced~\cite{Glendenning1965}.

%For the DWBA analysis, we chose appropriate optical model parameters (OMPs) for the incoming $d+^{138}$Ba channel by comparing our measured elastic scattering angular distribution with DWBA results from using different global OMPs.  This comparison showed that the recommended OMPs by An and Cai~\cite{AnCai2006} yielded best agreement with our data. For the outgoing $\alpha$+$^{136}$Cs channel we chose the OMPs of  Burnett \textit{et al.}~\cite{Burnett1985}, as they were optimized for the \isotope[136]Ba$(\alpha,\alpha)$ reaction at 20~MeV~\footnote{The maximum energy of the outgoing $\alpha$'s in our \isotope[138]Ba$(d,\alpha)$ reaction is not markedly different, at approximately 30~MeV.}.
For the DWBA analysis, we chose appropriate optical model parameters (OMPs) for the incoming $d+^{138}$Ba channel~\cite{AnCai2006} by comparing our measured elastic scattering angular distribution with DWBA results from using different global OMPs. % This comparison showed that the recommended OMPs by An and Cai~\cite{AnCai2006} yielded best agreement with our data.
For the outgoing $\alpha$+$^{136}$Cs channel we chose the OMPs of  Ref.~\cite{Burnett1985}, which were optimized for the \isotope[136]Ba$(\alpha,\alpha)$ reaction at 20~MeV~\footnote{The maximum energy of the outgoing $\alpha$'s in our \isotope[138]Ba$(d,\alpha)$ reaction is not markedly different, at approximately 30~MeV.}.
\begin{figure*}[t]
  \centering
  \includegraphics[scale=0.39]{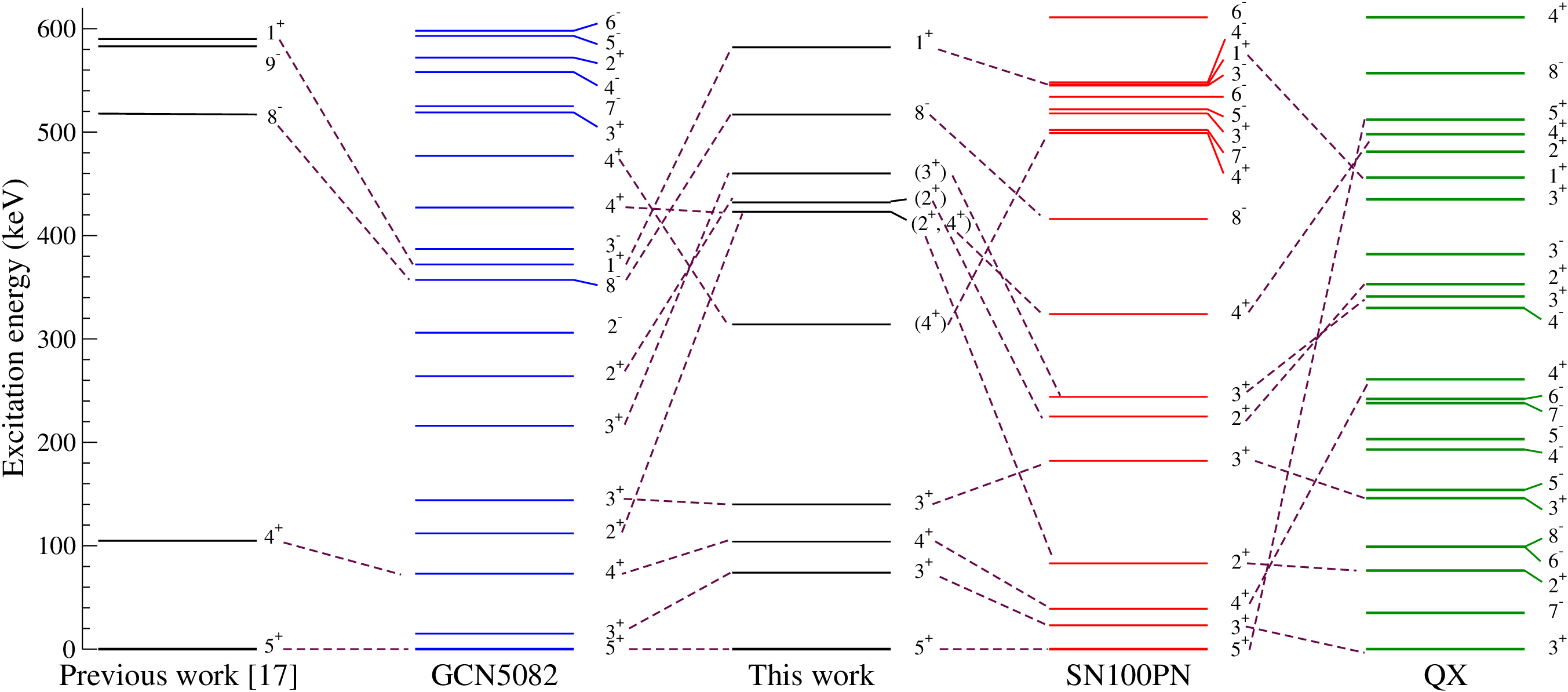} \\
  \caption{Comparison between theory and experiment for the low-lying energy spectrum of \isotope[136]Cs. The shell-model results were obtained with the GCN5082, SN100PN and QX effective interactions.}
  \label{Fig:levels} 
  \end{figure*}
The \isotope[138]Ba$(d,\alpha)$ calculations were performed assuming the `cluster' deuteron-transfer approximation~\cite{Curry,Rob1}, with form-factors for a deuteron in a Woods-Saxon potential well, at the correct separation energy for each state in $^{136}$Cs. %On account of the large angular momentum mismatch between the incoming deuteron and the outgoing $\alpha$,
We also took into consideration finite-range corrections~\cite{DaehnickPark1969,Charlton} and nonlocality effects, using the prescription from
Ref.~\cite{DelVecchio1972}. Next, our measured cross section angular distributions were overlaid with normalized best-fit DWBA results. The latter were obtained assuming various $L$-transfer values for given $J$, and allowed incoherent summations of two different values $L$ and $L'$. Identified states were then compared with shell-model predictions and previous measurements.

For the shell-model calculations we used a configuration space comprising the $0g_{7/2},~1d_{5/2},~1d_{3/2},~2s_{1/2}$ and $0h_{11/2}$ orbitals for neutrons and protons, and three different Hamiltonians: SN100PN~\cite{sn100}, GCN5082~\cite{gcn} and QX~\cite{qx}. The SN100PN interaction is very similar to the JJ55t Hamiltonian~\cite{Rebeiro:20}, and was used by Ref.~\cite{Scott:20} to evaluate the level scheme of \isotope[136]Cs.
Independently, the GCN5082 and QX Hamiltonians were used to calculate the \isotope[136]Xe $0\nu2\beta$ NME~\cite{javier,Neascu:15}.
%and $2\nu2\beta$ NME~\cite{Gando:19}.
%
%
\begin{table}[t]
  \begin{flushleft}
% \begin{adjustbox}{width=1.0\textwidth}
  \caption{\label{tab:136Cs_Energies}
 Observed \isotope[136]Cs levels up to the $1_1^+$ state.
 %Complete results for all states observed in this work (up to 2.5~MeV) are presented in the supplemental material. 
  }
  \begin{ruledtabular}
  \begin{tabular}{cclccc}
  \multicolumn{2}{c }{Refs.~\cite{Mccutchan2018,Frekers2013}} &  \multicolumn{4}{c}{This work}  \\
  \cline{1-2}\cline{3-6}
  \multicolumn{1}{l}{$E_x$ (keV)} &  \multicolumn{1}{c}{$J^{\pi}$} & \multicolumn{1}{l}{$E_x$ (keV)} &  \multicolumn{1}{c}{$L$} & \multicolumn{1}{c}{$L'$}& \multicolumn{1}{c}{Assigned $J^\pi$}  \\
  %\multicolumn{1}{c}{(keV)} &   & \multicolumn{1}{c}{(keV)} &  &  &  \\
  \colrule  
  0.0    &$5^+$ &0.0     & 4 & 6   &$5^+$       \\
         &      &74(2)   & 4   & ... &$3^+$  \\
104.8(3) &$4^+$ &104(2)$^a$  & 4   & ... &$4^+$  \\
         &      &140(3)  & 2   & 4   &$3^+$ \\
         &      &314(2)  & 4   & ... &$(4^+)$     \\
         &      &423(3)$^b$  & 4   & ... &$(4^+)$\\
431(2)   &$(3^+)$&432(3) &2    & ... &$(2^+)$\\
         &      &460(3)  &4    &...&$(3^+)$\\
517.9(1) &$8^-$ &517(3)  & 7   & 9   &$8^-$ \\
583.9(5) &$9^-$ &...     &...  &...  &...\\
591(2)   &$1^+$ &589(3)  & 0   &2    &$1^+$
  %\hline 
  %\hline
  \end{tabular}
$^a$~Although the measured angular distribution for this state is dissimilar to other $L = 4$ cases, our spin-parity assignment is consistent with a previous $\gamma$-ray measurement~\cite{Wimmer:11}.\\
$^b$~Possible unresolved $(4^+)$, $(2^+)$ doublet. See text for details.
\end{ruledtabular}
\end{flushleft}
 \end{table}

\begin{figure}[t]
  \includegraphics[scale=0.16]{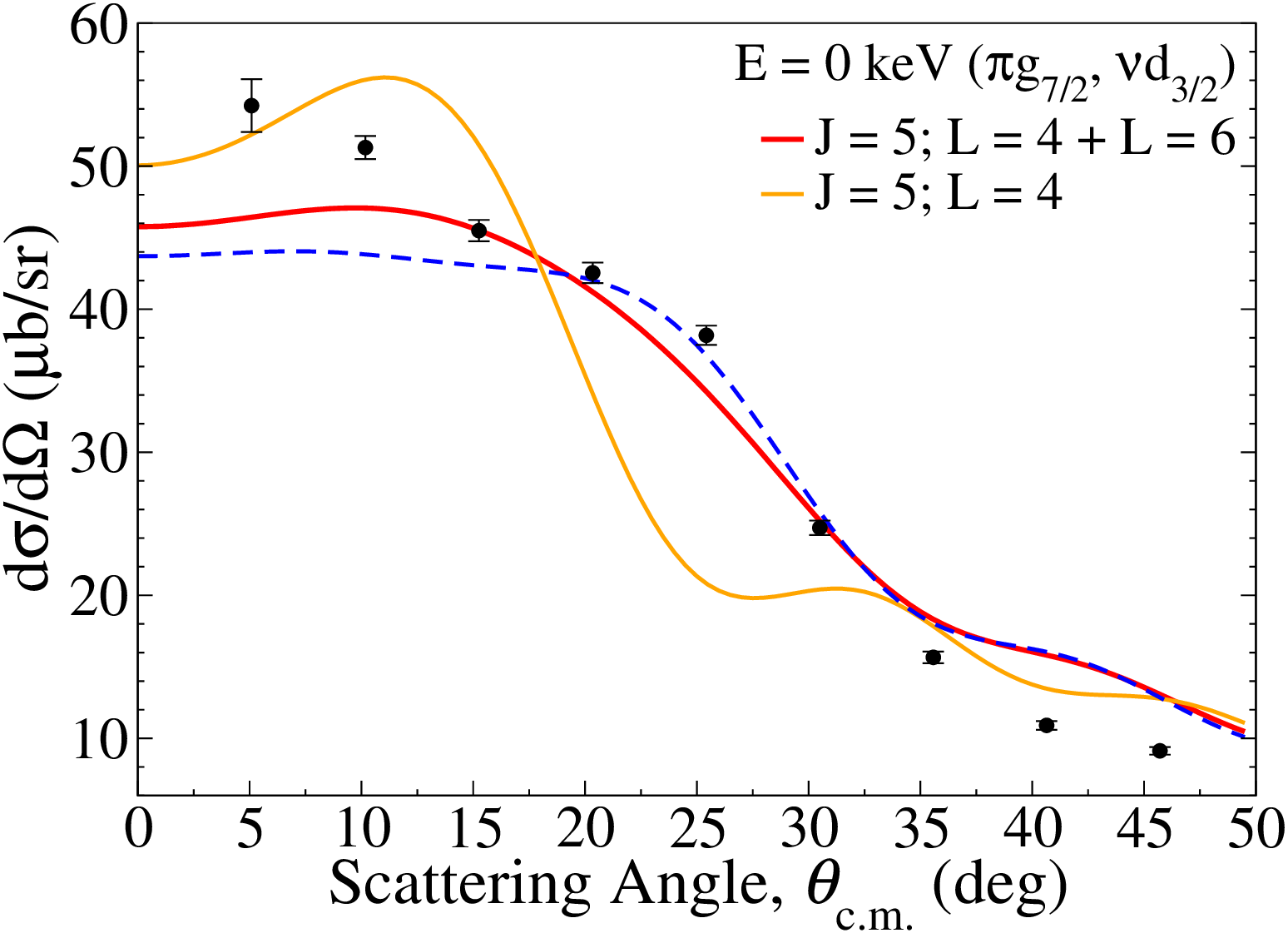}
  \includegraphics[scale=0.16]{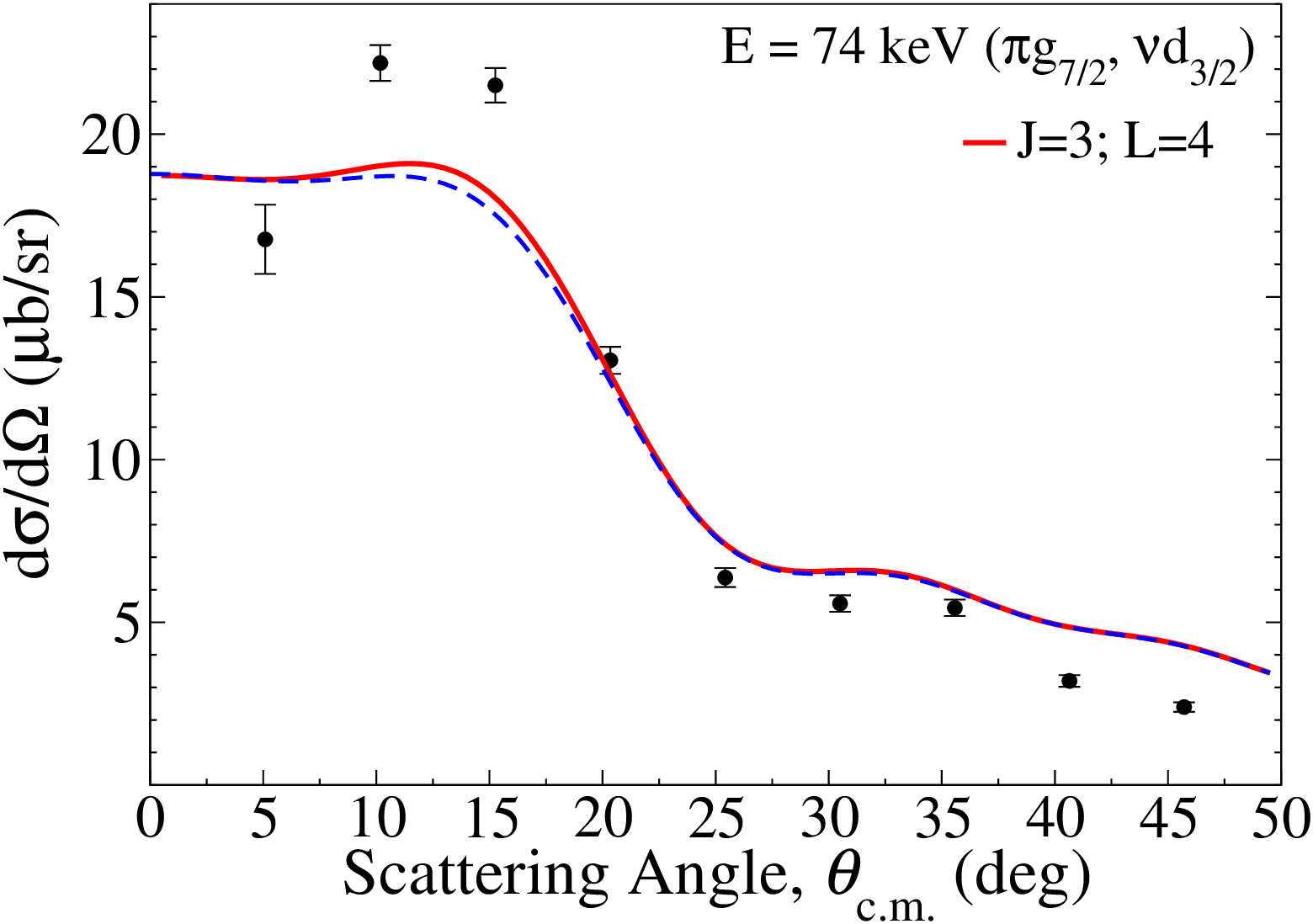}
  \includegraphics[scale=0.16]{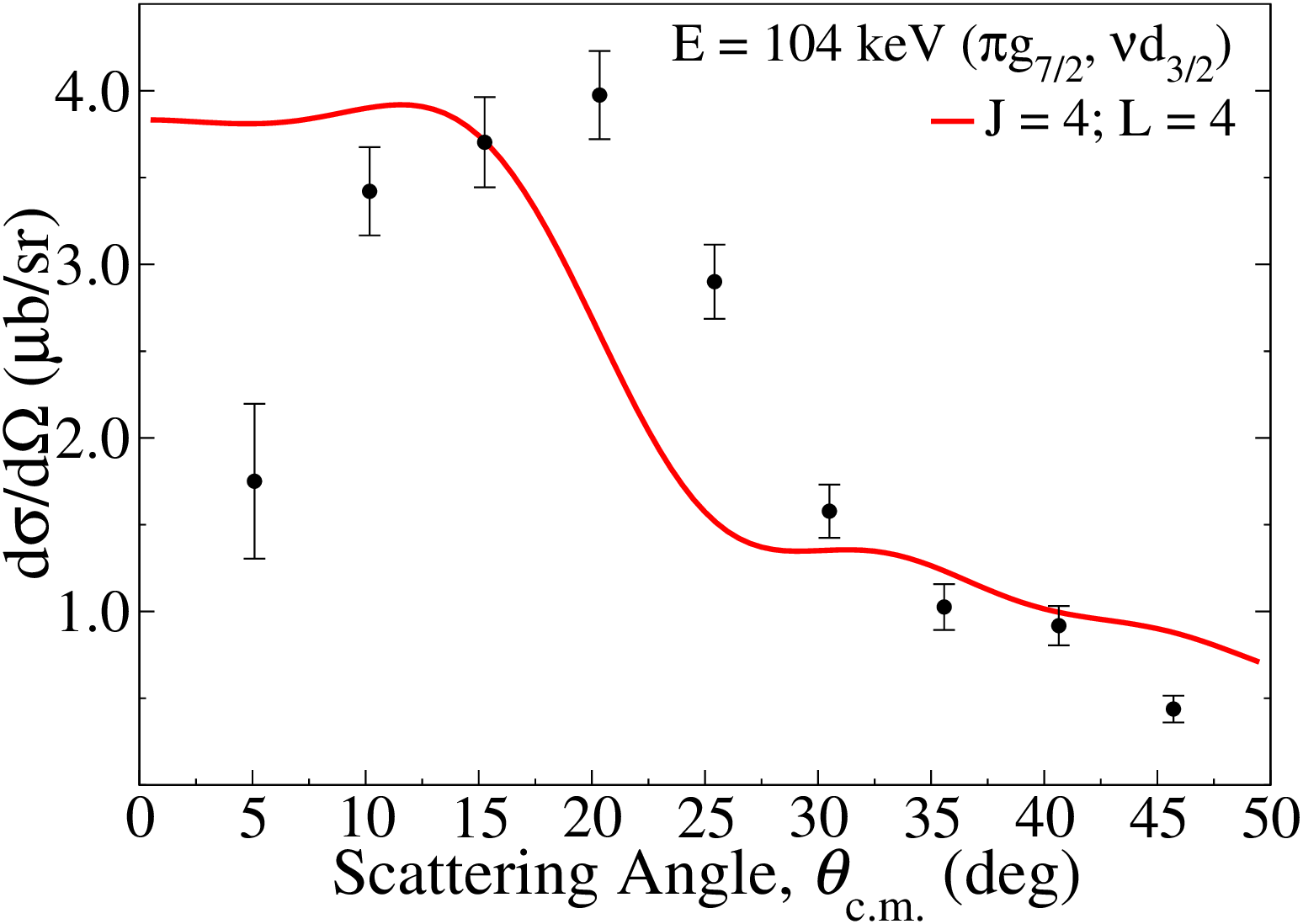}
  \includegraphics[scale=0.16]{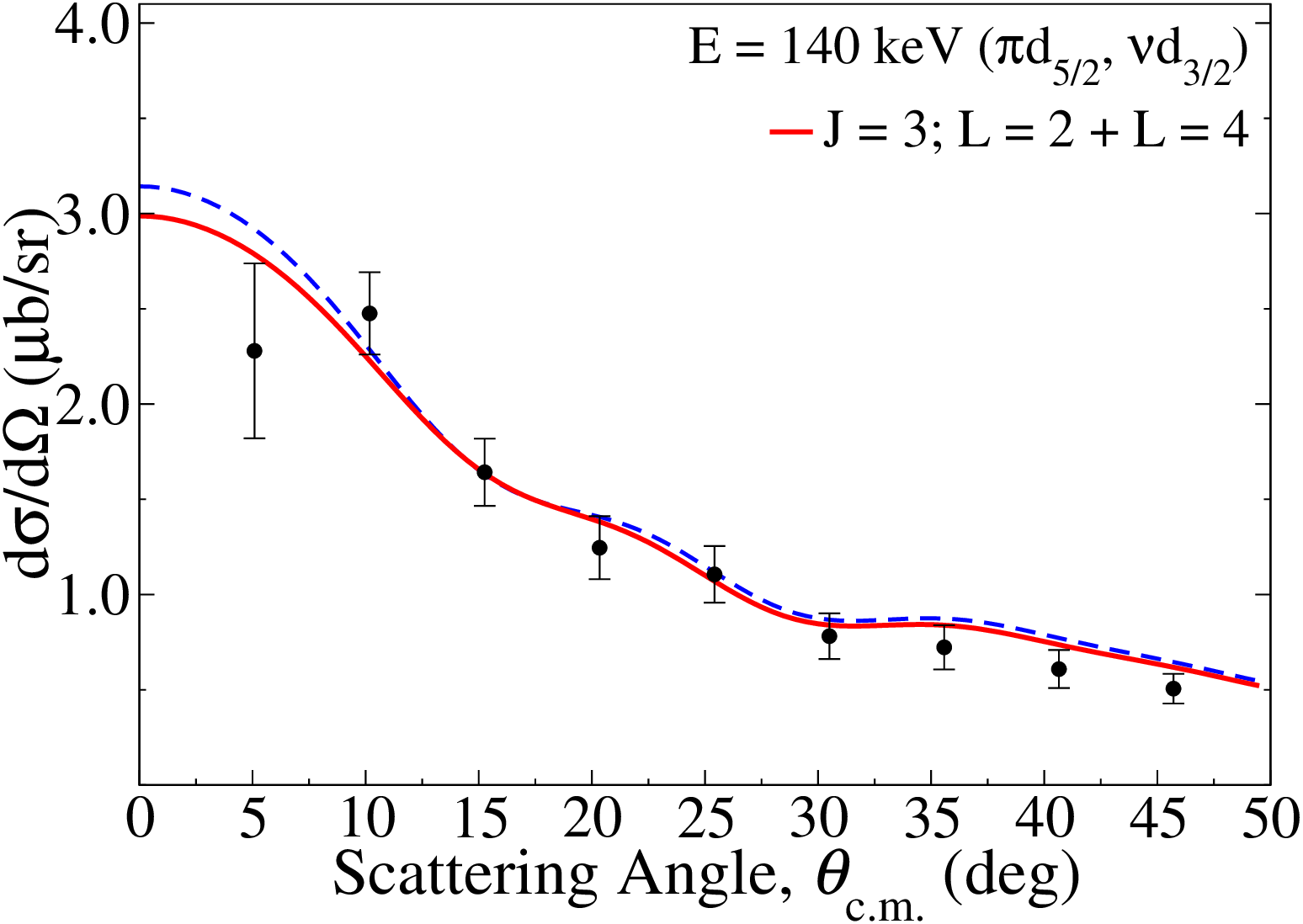}
  \includegraphics[scale=0.16]{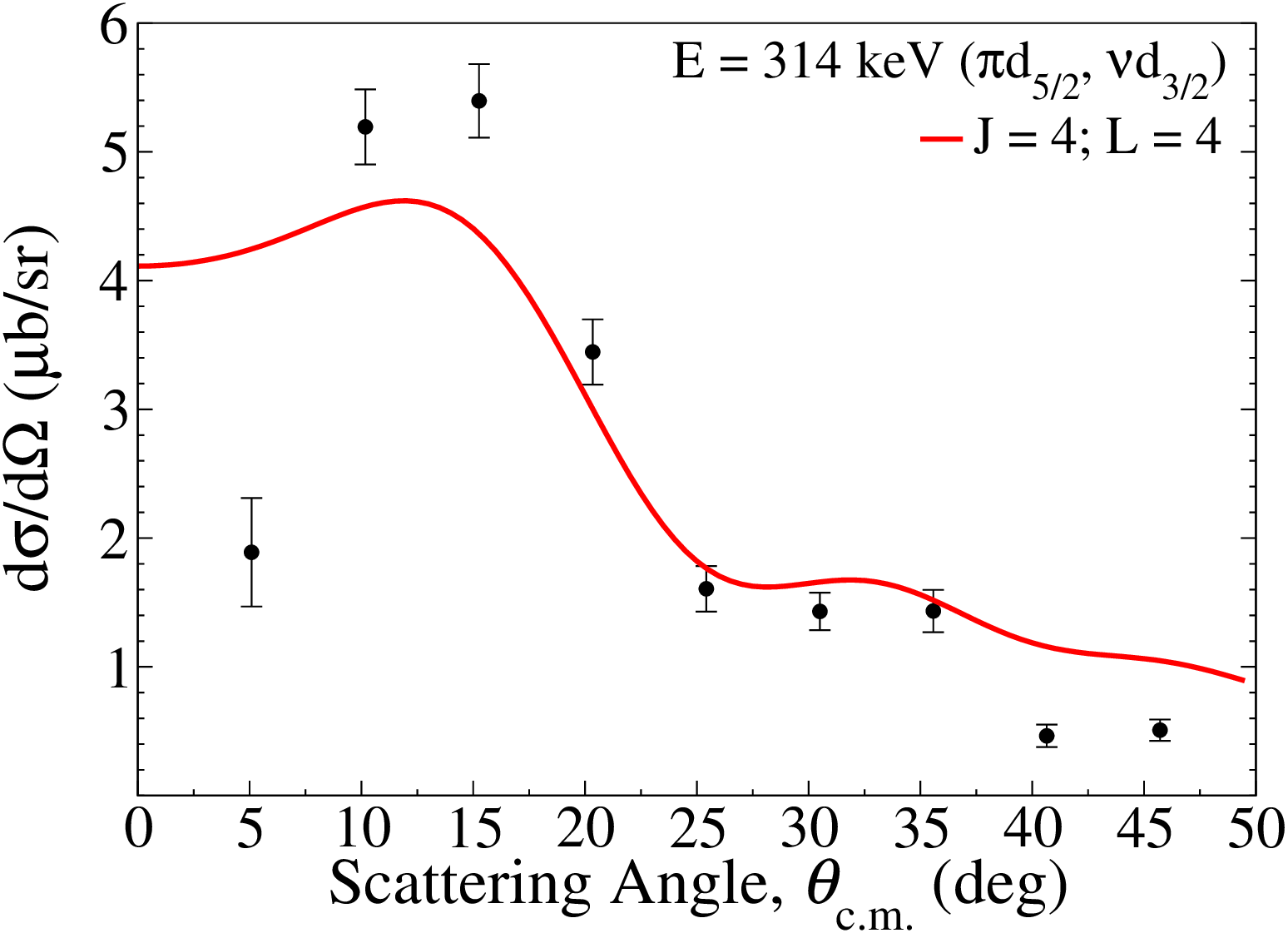}
  \includegraphics[scale=0.16]{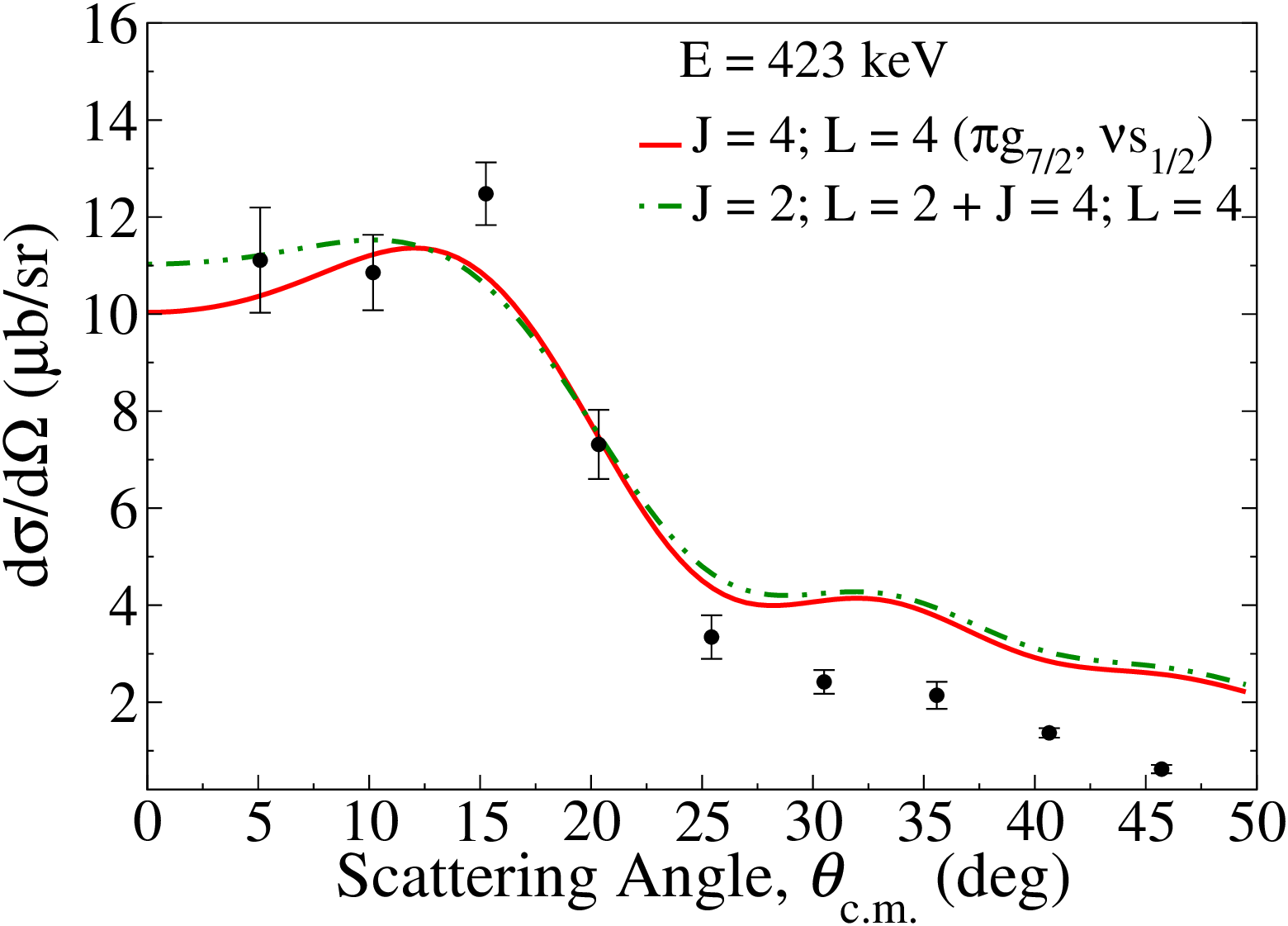}
    \includegraphics[scale=0.16]{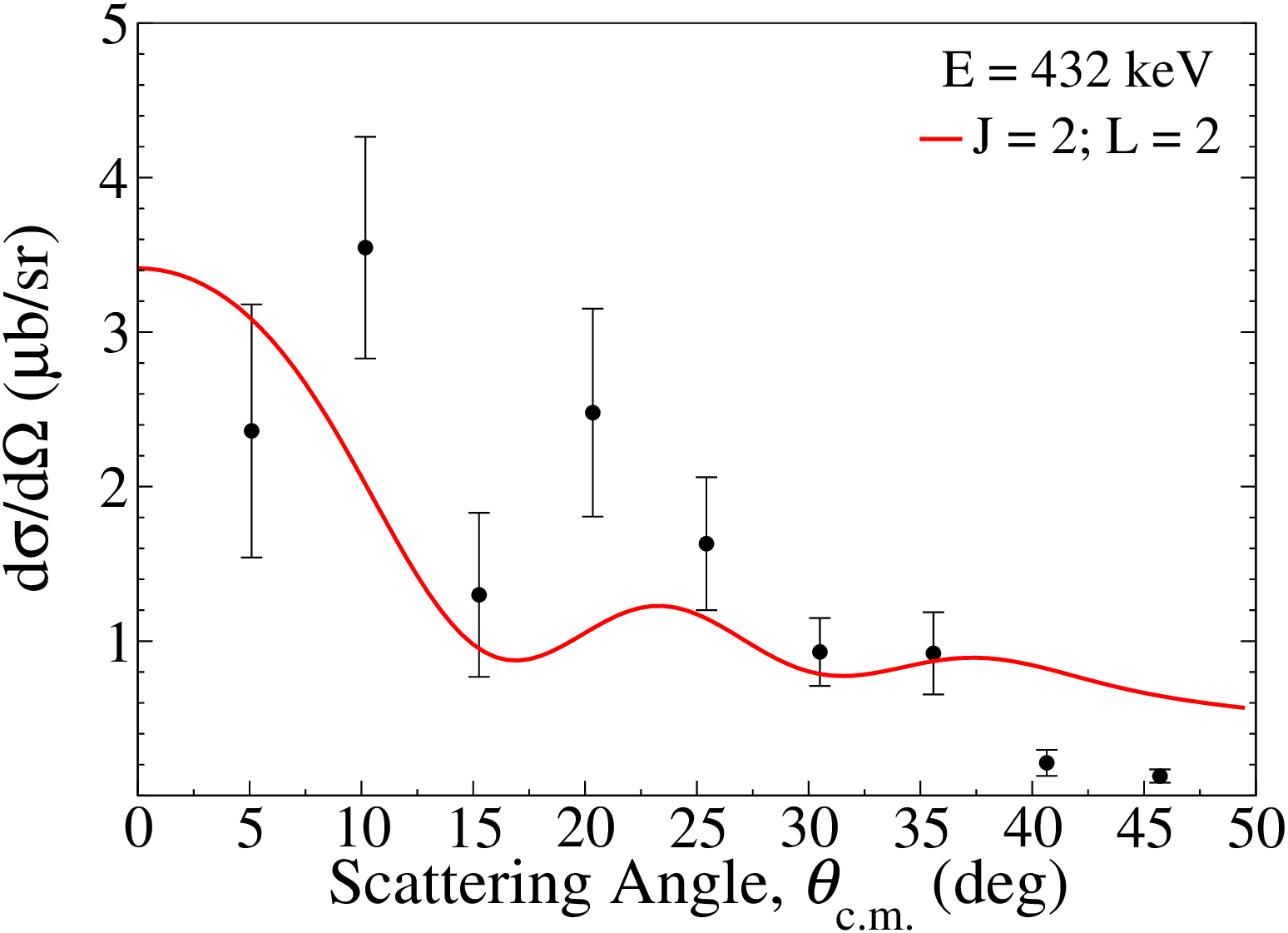}
  \includegraphics[scale=0.16]{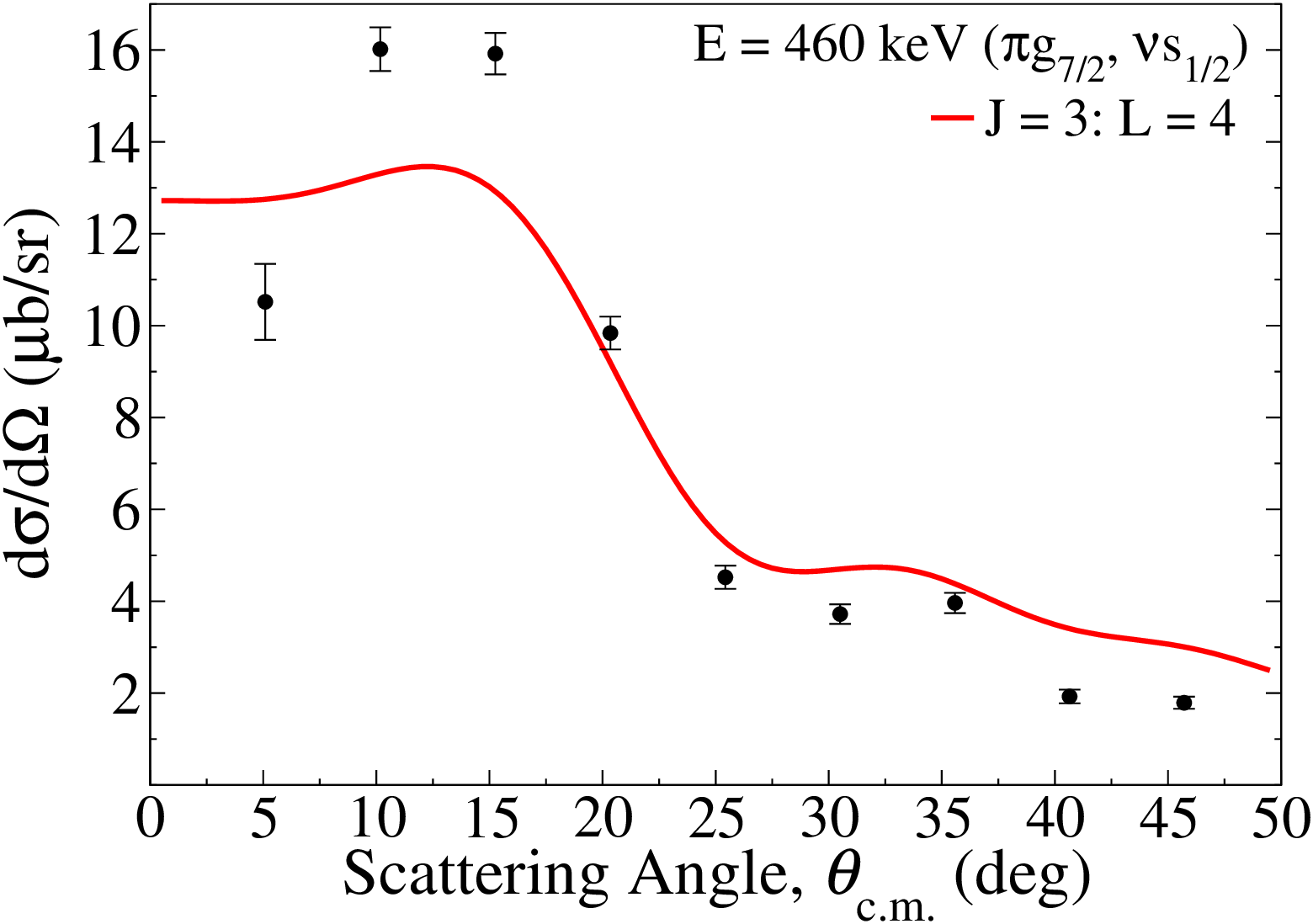}
  \includegraphics[scale=0.16]{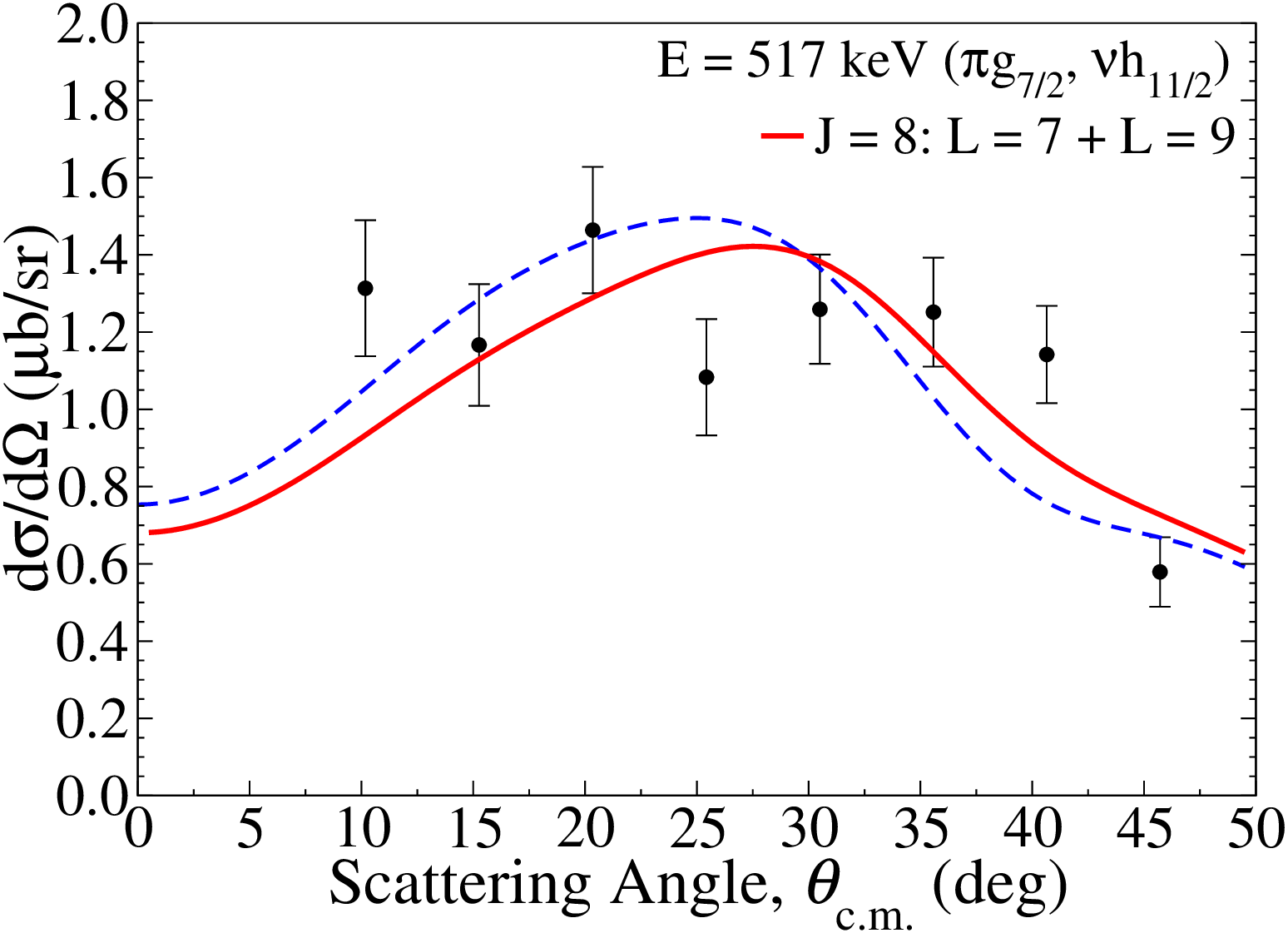}
% \hspace*{-10.8 em}
 % \includegraphics[scale=0.16]{136Cs_589_final.eps}
  \includegraphics[scale=0.16]{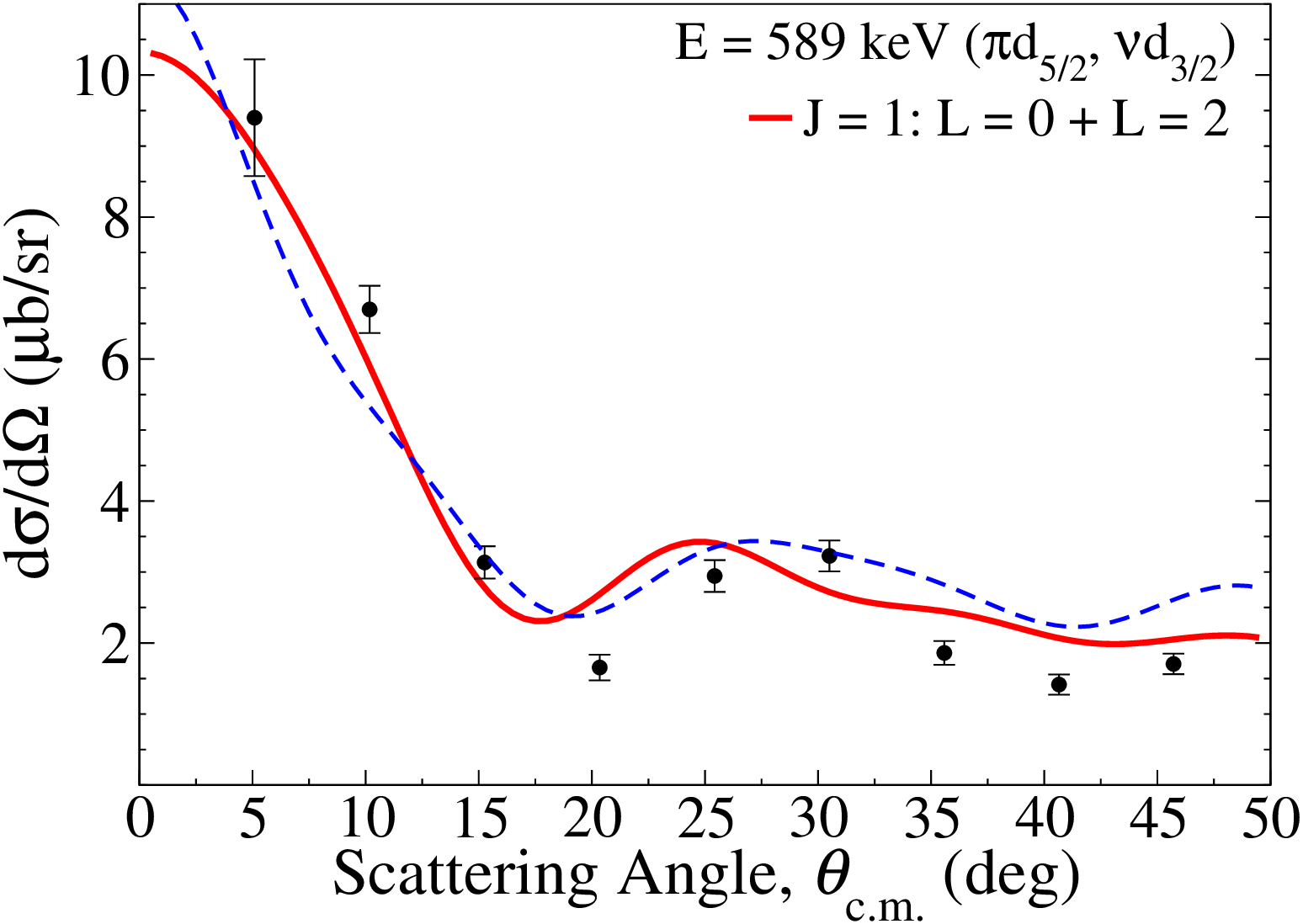}
  \caption{\label{Fig:angdists1}Measured \isotope[138]Ba$(d,\alpha)$ angular distributions compared with best-fit DWUCK5 DWBA predictions (solid red curves). The blue dashed curves are from using fixed relative $L$ contributions from Eq.~\eqref{eq:9j}. The dominant orbitals involved in the pair-transfer are specified in each plot.}
\end{figure}

Figure~\ref{Fig:levels} compares calculated energy levels of \isotope[136]Cs to those identified from this experiment. Our results, for states up to the $1_1^+$ level are summarized in Fig.~\ref{Fig:angdists1} and Table~\ref{tab:136Cs_Energies}. We also used two-nucleon transfer amplitudes (TNAs)~\cite{Brown:14} obtained with the GCN5082 and SN100PN Hamiltonians for critical comparative cross-checks. This was feasible because most of the low-lying states had TNA dominated by simple two-nucleon configurations. For example, both calculations showed that the dominant orbitals involved in the transfer to the $J ^\pi = 5^+$ ground state~\cite{Dabb:71,Thibault:81} are
$g_{7/2}$ and $d_{3/2}$ for proton $(\pi)$ and neutron $(\nu)$ pick-up, respectively. This state can be produced by both $L = 4$ and $L = 6$ transfer.
%If one ignores admixtures of other configurations and sequential/multi-step transfer,
The relative $L$ contributions can be evaluated via the $jj$ to $LS$ transformation that involves the normalized $9j$ coefficient~\cite{Glendenning:63},
\begin{equation}
\sqrt{3(2j_n+1)(2j_p+1)(2L+1)}
   \begin{Bmatrix} 
  l_n & 1/2 & j_n \\
  l_p & 1/2 & j_p \\
  L   & 1   & J
   \end{Bmatrix}.
   \label{eq:9j}
\end{equation}
This yields a predominantly $L = 6$ transition for the ground state, which is consistent with our observations. The same two-nucleon configuration dominates transfer to the $3_1^+$ and $4_1^+$ states. For the former, the intensity of the $L = 2$ transition is nearly 17 times weaker than $L = 4$ transfer. This agrees with the measured angular distribution of the first excited state, observed at 74~keV. Next, we compared the measured cross section for this level relative to the ground state (after accounting for the difference in their predicted DWBA yields), with the relative scaling of their calculated transfer intensities. The reasonable agreement between these two values validated the $3_1^+$ assignment for this state.
In comparison, we identify the 140-keV state as $3_2^+$, whose dominant TNA corresponds to the $(\pi d_{5/2})~(\nu d_{3/2})$ orbitals. Both $L = 2$ and $L = 4$ transfer contribute for this state, which
%with the former intensity being $\sim 2.4$~times the latter.
agrees well with the measured distribution.
Spin-parity assignments for the remaining states identified in Table~\ref{tab:136Cs_Energies} were made through similar analysis of the shapes of the angular distributions, relative cross sections, and $L$-transfer intensities predicted by theory.

We do not observe the explicit signatures of the low-lying $2^+$ states, which are predicted to be weakly populated. We also do not observe the known $9^-$ state at 583.9(5)~keV. This can be explained by the DWBA calculations, which show that $L = 9$ transfer for this state is significantly weaker than the dominant $L = 7$ transfer to the $8^-$ state.
A tentative $3^+$ state was reported at 431~keV~\cite{Frekers2013}, but excluded from Ref.~\cite{Mccutchan2018}'s compilation.
We investigated this state's possible existence by refitting the 423~keV peak with fixed lineshape parameters, based on previous knowledge of the detector response~\cite{Kamil:22}.
This analysis indicated a possible level at $E_\mathrm{x} = $~432(3)~keV, whose angular distribution is shown in Fig.~\ref{Fig:angdists1}. Although it is statistics-limited, the measured distribution appears to be consistent with $L = 2$ transfer.
The intensity of this possible transition is comparable to those predicted for the $2_1^+$ and $2_2^+$ levels.
%A comparison of measured cross sections with relative intensities obtained with the TNA show that this  state is a better match for either of the predicted $2_1^+/2_2^+$ levels.
%
% Unlike the latter $(L = 4)$, their measured angular distributions are consistent with $L = 2$ transfer. The 432~keV state agrees with a previously reported $(3^+)$ level~\cite{Frekers2013}. We identify this to be the $3_4^+$ state, whose TNA are primarily described by the  configuration. Since the corresponding $L = 4$ 9-$j$ coefficient for this configuration is zero, its measured angular distribution is expected to have only an $L = 2$ component. This agrees reasonably well with the data in Fig~\ref{Fig:angdists1}.
We also observe that the $\theta_{\rm lab} = 5^\circ$ cross section for the 423~keV state is enhanced compared to the other $L = 4$ transitions.
This can be attributed to an additional $L = 2$ component which is $\sim 20\%$ of the $L = 4$ contribution, as shown in Fig.~\ref{Fig:angdists1}.
%This can be described by an additional $L = 2$ component (which is $\sim 20\%$ of the $L = 4$ contribution, from the relative TNA of either of the $2_1^+/2_2^+$ levels), as shown in Fig.~\ref{Fig:angdists1}.
Thus, one cannot rule out an unresolved state at $\sim 423$~keV,  with an $L = 2$ contribution that corresponds to one of the $2^+$ levels.
%
% The former was listed as a tentative $(3^+)$
% These aspects are also discussed in the supplemental file.

%The dominant $(\pi g_{7/2})$, $(\pi d_{5/2})$, $(\nu d_{3/2})$, $(\nu s_{1/2})$ and $(\nu h_{11/2})$ configurations that describe these levels are also consistent with other observations. The
%spin-parity values for first two (three) states in \isotope[137]Cs $(N = 82, Z = 55)$ and \isotope[137]Ba $(N = 81, Z = 56)$, are $7/2^+$, $5/2^+$, and $3/2^+$, $1/2^+$, $11/2^-$, respectively~\cite{Mccutchan2018}. This information, together with independently measured \isotope[138]Ba$(d,\isotope[3]He)$ spectroscopic factors~\cite{Wildenthal}, provides an important validation of these configurations.
   
%
%The results for the \isotope[136]Cs energy spectrum are shown in
%The high suppression of the quadrupole contribution is apparent for almost all of our assigned $3^+$ states in Table~\ref{tab:136Cs_Energies}. These observations were validated by  The TNAs were dominated by $L = 4$ transitions for all low-lying $J^\pi = 3^+$ levels, apart from the $3_2^+$ state which had a competing $L = 2$ component.
%Future higher-resolution experimental work to identify the predicted $2^+$ states will be useful in this regard.

%
Figure~\ref{Fig:levels} shows that the SN100PN and GCN5082 results are overall very similar and could be matched to our identified levels from this experiment. A recent independent calculation performed with the proton–neutron quasiparticle random-phase approximation (pnQRPA)~\cite{Gimeno:23} also shows reasonable overall agreement with our measured spectrum.
However, there is a stark disagreement with the QX results where the $5_1^+$ ground state shows up at a significantly higher energy. The QX interaction also shows several low-lying negative parity states in \isotope[136]Cs that are not predicted by the other Hamiltonians or verified by experiment.
%These observations are also consistent with  recent independent calculations performed with the proton–
%neutron quasiparticle random-phase approximation .
These observations underscore the importance of testing model predictions in intermediate odd-odd nuclei for $0\nu2\beta$ candidates. Under such requirement, the QX interaction may be considered less reliable and likewise disfavor \isotope[136]Xe $0\nu 2\beta$ NME values determined with this Hamiltonian.

In the context of solar $\nu_e$/FDM detection in xenon-based detectors, this work presents the ﬁrst unequivocal identification of the predicted long-lived excited $3_1^+$ state in \isotope[136]{Cs}, with a ﬁrm spin-parity assignment. The measured excitation energy, $E_\mathrm{x}$ = 74~keV, is more than three times higher than the shell-model prediction in Ref.~\cite{Scott:20}.  In the absence of competing branches~\footnote{A careful analysis of our $\alpha$ spectrum rules out any possible intermediate $4^+$ state below 74~keV, which would enable faster $M1$ transitions and lead to a substantially shorter lifetime for the $3_1^+$ level.}, the $3_1^+$ state at 74~keV is expected to still have a long enough lifetime for a feasible delayed coincidence tagging of solar $\nu_e$/FDM interactions on \isotope[136]{Xe}. As this level can deexcite to the $5_1^+$ ground state via both internal conversion (IC) and $\gamma$-ray emission, its total transition rate is proportional to $E_\gamma^5(1+\alpha)$, where $\alpha$ is the IC coefficient~\cite{ic}. Based on our measured energy and simple scaling arguments, the shell-model predicted lifetime of the state is $\sim 280$~ns, three times shorter than the value obtained with $E_\mathrm{x} = 23$~keV~\cite{Scott:20,Note1}.

In conclusion, we used \isotope[138]{Ba}$(d,\alpha)$ angular distribution measurements, together with shell-model calculations to report the location of possible metastable states with $J \ge 1$ in the odd-odd \isotope[136]{Cs} nucleus. The observed new states offer an opportunity for high background rejection, and open new possibilities for the detection of solar $\nu_e$ events and/or FDM interactions in large xenon detectors.
We unambiguously identify the first excited state in \isotope[136]{Cs}, which is has spin-parity $3^+$ and would decay to the $5^+$ ground state via a slow $E2$ transition. Our findings are supported by a recent independent study~\cite{tunl} that measured the lifetime of the $3_1^+$ state to be $\tau = 157(4)$~ns. %The measurements reported in Ref.~\cite{tunl} and our work complement each other well, and emphasize the importance of experimental observables in $^{136}$Cs for new physics searches.
We also compare our experimental results with shell-model predictions made with three Hamiltonians that were previously used to evaluate the \isotope[136]{Xe} $0 \nu 2 \beta$ NME. The comparison shows that one of the Hamiltonians (QX), which also systematically underestimates the NME compared to the others, fails to accurately describe the \isotope[136]{Cs} spectrum. As this inadequacy may have been obscured when predictions were compared with experimental data on even-even nuclei, one might disfavor $0 \nu 2 \beta$ results obtained with this
Hamiltonian.

%mainly because of the dominant pairing interaction and other collective properties.

 \begin{acknowledgments}
This work was partially supported by the National Research Foundation (NRF), South Africa, under Grant No. 85100, the Natural Sciences and Engineering Research Council of Canada (NSERC), the U.S. National Science Foundation under Grant No. PHY-2110365 and the U.S. Department of Energy (DOE) under Grant No. DE-FG02-93ER40789.
\end{acknowledgments}

%This result calls for 
% 
% of 
%and further experimental studies of $\gamma$-ray transitions connecting the low-lying states within
%\isotope[136]Cs 
%it is apparent from the previous analysis of Ref.~\cite{Scott:20} that a lifetime of $\mathcal{O}(1)~\mu$s should be adequate to identify such events.   
% 
% In summary we performed a high-resolution \isotope[138]Ba$(d,\alpha)$ study of states in the odd-odd \isotope[136]Cs, the intermediate nucleus in \isotope[136]Xe $2\beta$ decay. Our observations for the (until now) largely unexplored energy region below the $1_1^+$ state are compared to calculations performed with three shell-model Hamiltonians that were used to evaluate critical NMEs related to $2\beta$ decay~\cite{Caurier:08,Rebeiro:20,Romeo:22,Gando:19,Lotta:22}. This comparison calls into question one of the Hamiltonians that dramatically fails to describe the experimentally observed spectrum. We also show that the first excited $3^+$ state in \isotope[136]Cs is more than three times higher than calculated values.

%The large level density at higher energies prevents us from extending such a comparison to higher-lying states.  

%\section{\label{sec:acknowledgment}Acknowledgments}
%   
\bibliography{138Ba_da_final_version.bib}        %use a bibtex bibliography file refs.bib

\begin{thebibliography}{57}%
\makeatletter
\providecommand \@ifxundefined [1]{%
 \@ifx{#1\undefined}
}%
\providecommand \@ifnum [1]{%
 \ifnum #1\expandafter \@firstoftwo
 \else \expandafter \@secondoftwo
 \fi
}%
\providecommand \@ifx [1]{%
 \ifx #1\expandafter \@firstoftwo
 \else \expandafter \@secondoftwo
 \fi
}%
\providecommand \natexlab [1]{#1}%
\providecommand \enquote  [1]{``#1''}%
\providecommand \bibnamefont  [1]{#1}%
\providecommand \bibfnamefont [1]{#1}%
\providecommand \citenamefont [1]{#1}%
\providecommand \href@noop [0]{\@secondoftwo}%
\providecommand \href [0]{\begingroup \@sanitize@url \@href}%
\providecommand \@href[1]{\@@startlink{#1}\@@href}%
\providecommand \@@href[1]{\endgroup#1\@@endlink}%
\providecommand \@sanitize@url [0]{\catcode `\\12\catcode `\$12\catcode
  `\&12\catcode `\#12\catcode `\^12\catcode `\_12\catcode `\%12\relax}%
\providecommand \@@startlink[1]{}%
\providecommand \@@endlink[0]{}%
\providecommand \url  [0]{\begingroup\@sanitize@url \@url }%
\providecommand \@url [1]{\endgroup\@href {#1}{\urlprefix }}%
\providecommand \urlprefix  [0]{URL }%
\providecommand \Eprint [0]{\href }%
\providecommand \doibase [0]{http://dx.doi.org/}%
\providecommand \selectlanguage [0]{\@gobble}%
\providecommand \bibinfo  [0]{\@secondoftwo}%
\providecommand \bibfield  [0]{\@secondoftwo}%
\providecommand \translation [1]{[#1]}%
\providecommand \BibitemOpen [0]{}%
\providecommand \bibitemStop [0]{}%
\providecommand \bibitemNoStop [0]{.\EOS\space}%
\providecommand \EOS [0]{\spacefactor3000\relax}%
\providecommand \BibitemShut  [1]{\csname bibitem#1\endcsname}%
\let\auto@bib@innerbib\@empty
%</preamble>
\bibitem [{\citenamefont {Raghavan}(1997)}]{Raghavan}%
  \BibitemOpen
  \bibfield  {author} {\bibinfo {author} {\bibfnamefont {R.~S.}\ \bibnamefont
  {Raghavan}},\ }\href {\doibase 10.1103/PhysRevLett.78.3618} {\bibfield
  {journal} {\bibinfo  {journal} {Phys. Rev. Lett.}\ }\textbf {\bibinfo
  {volume} {78}},\ \bibinfo {pages} {3618} (\bibinfo {year}
  {1997})}\BibitemShut {NoStop}%
\bibitem [{\citenamefont {Adhikari}\ \emph {et~al.}(2021)\citenamefont
  {Adhikari}, \citenamefont {Kharusi}, \citenamefont {Angelico}, \citenamefont
  {Anton}, \citenamefont {Arnquist}, \citenamefont {Badhrees}, \citenamefont
  {Bane}, \citenamefont {Belov}, \citenamefont {Bernard}, \citenamefont
  {Bhatta} \emph {et~al.}}]{Adhikari:21}%
  \BibitemOpen
  \bibfield  {author} {\bibinfo {author} {\bibfnamefont {G.}~\bibnamefont
  {Adhikari}}, \bibinfo {author} {\bibfnamefont {S.~A.}\ \bibnamefont
  {Kharusi}}, \bibinfo {author} {\bibfnamefont {E.}~\bibnamefont {Angelico}},
  \bibinfo {author} {\bibfnamefont {G.}~\bibnamefont {Anton}}, \bibinfo
  {author} {\bibfnamefont {I.~J.}\ \bibnamefont {Arnquist}}, \bibinfo {author}
  {\bibfnamefont {I.}~\bibnamefont {Badhrees}}, \bibinfo {author}
  {\bibfnamefont {J.}~\bibnamefont {Bane}}, \bibinfo {author} {\bibfnamefont
  {V.}~\bibnamefont {Belov}}, \bibinfo {author} {\bibfnamefont {E.~P.}\
  \bibnamefont {Bernard}}, \bibinfo {author} {\bibfnamefont {T.}~\bibnamefont
  {Bhatta}},  \emph {et~al.},\ }\href {\doibase 10.1088/1361-6471/ac3631}
  {\bibfield  {journal} {\bibinfo  {journal} {J. Phys. G}\ }\textbf {\bibinfo
  {volume} {49}},\ \bibinfo {pages} {015104} (\bibinfo {year}
  {2021})}\BibitemShut {NoStop}%
\bibitem [{\citenamefont {Abe}\ \emph {et~al.}(2023)\citenamefont {Abe},
  \citenamefont {Asami}, \citenamefont {Eizuka}, \citenamefont {Futagi},
  \citenamefont {Gando}, \citenamefont {Gando}, \citenamefont {Gima},
  \citenamefont {Goto}, \citenamefont {Hachiya}, \citenamefont {Hata} \emph
  {et~al.}}]{kamland}%
  \BibitemOpen
  \bibfield  {author} {\bibinfo {author} {\bibfnamefont {S.}~\bibnamefont
  {Abe}}, \bibinfo {author} {\bibfnamefont {S.}~\bibnamefont {Asami}}, \bibinfo
  {author} {\bibfnamefont {M.}~\bibnamefont {Eizuka}}, \bibinfo {author}
  {\bibfnamefont {S.}~\bibnamefont {Futagi}}, \bibinfo {author} {\bibfnamefont
  {A.}~\bibnamefont {Gando}}, \bibinfo {author} {\bibfnamefont
  {Y.}~\bibnamefont {Gando}}, \bibinfo {author} {\bibfnamefont
  {T.}~\bibnamefont {Gima}}, \bibinfo {author} {\bibfnamefont {A.}~\bibnamefont
  {Goto}}, \bibinfo {author} {\bibfnamefont {T.}~\bibnamefont {Hachiya}},
  \bibinfo {author} {\bibfnamefont {K.}~\bibnamefont {Hata}},  \emph {et~al.}
  (\bibinfo {collaboration} {KamLAND-Zen Collaboration}),\ }\href {\doibase
  10.1103/PhysRevLett.130.051801} {\bibfield  {journal} {\bibinfo  {journal}
  {Phys. Rev. Lett.}\ }\textbf {\bibinfo {volume} {130}},\ \bibinfo {pages}
  {051801} (\bibinfo {year} {2023})}\BibitemShut {NoStop}%
\bibitem [{\citenamefont {Adams}\ \emph {et~al.}(2021)\citenamefont {Adams},
  \citenamefont {{\'A}lvarez}, \citenamefont {Arazi}, \citenamefont {Arnquist},
  \citenamefont {Azevedo}, \citenamefont {Bailey}, \citenamefont {Ballester},
  \citenamefont {Benlloch-Rodr{\'i}guez}, \citenamefont {Borges}, \citenamefont
  {Byrnes}, \citenamefont {C{\'a}rcel}, \citenamefont {Carri{\'o}n} \emph
  {et~al.}}]{next}%
  \BibitemOpen
  \bibfield  {author} {\bibinfo {author} {\bibfnamefont {C.}~\bibnamefont
  {Adams}}, \bibinfo {author} {\bibfnamefont {V.}~\bibnamefont {{\'A}lvarez}},
  \bibinfo {author} {\bibfnamefont {L.}~\bibnamefont {Arazi}}, \bibinfo
  {author} {\bibfnamefont {I.~J.}\ \bibnamefont {Arnquist}}, \bibinfo {author}
  {\bibfnamefont {C.~D.~R.}\ \bibnamefont {Azevedo}}, \bibinfo {author}
  {\bibfnamefont {K.}~\bibnamefont {Bailey}}, \bibinfo {author} {\bibfnamefont
  {F.}~\bibnamefont {Ballester}}, \bibinfo {author} {\bibfnamefont {J.~M.}\
  \bibnamefont {Benlloch-Rodr{\'i}guez}}, \bibinfo {author} {\bibfnamefont
  {F.~I. G.~M.}\ \bibnamefont {Borges}}, \bibinfo {author} {\bibfnamefont
  {N.}~\bibnamefont {Byrnes}}, \bibinfo {author} {\bibfnamefont
  {S.}~\bibnamefont {C{\'a}rcel}}, \bibinfo {author} {\bibfnamefont {J.~V.}\
  \bibnamefont {Carri{\'o}n}},  \emph {et~al.},\ }\href {\doibase
  10.1007/JHEP08(2021)164} {\bibfield  {journal} {\bibinfo  {journal} {J. High
  Energy Phys.}\ }\textbf {\bibinfo {volume} {2021}},\ \bibinfo {pages} {164}
  (\bibinfo {year} {2021})}\BibitemShut {NoStop}%
\bibitem [{\citenamefont {Zhang}\ \emph {et~al.}(2022)\citenamefont {Zhang},
  \citenamefont {Abdukerim}, \citenamefont {Bo}, \citenamefont {Chen},
  \citenamefont {Chen}, \citenamefont {Chen}, \citenamefont {Cheng},
  \citenamefont {Cheng}, \citenamefont {Cui}, \citenamefont {Fan} \emph
  {et~al.}}]{pandax:22}%
  \BibitemOpen
  \bibfield  {author} {\bibinfo {author} {\bibfnamefont {D.}~\bibnamefont
  {Zhang}}, \bibinfo {author} {\bibfnamefont {A.}~\bibnamefont {Abdukerim}},
  \bibinfo {author} {\bibfnamefont {Z.}~\bibnamefont {Bo}}, \bibinfo {author}
  {\bibfnamefont {W.}~\bibnamefont {Chen}}, \bibinfo {author} {\bibfnamefont
  {X.}~\bibnamefont {Chen}}, \bibinfo {author} {\bibfnamefont {Y.}~\bibnamefont
  {Chen}}, \bibinfo {author} {\bibfnamefont {C.}~\bibnamefont {Cheng}},
  \bibinfo {author} {\bibfnamefont {Z.}~\bibnamefont {Cheng}}, \bibinfo
  {author} {\bibfnamefont {X.}~\bibnamefont {Cui}}, \bibinfo {author}
  {\bibfnamefont {Y.}~\bibnamefont {Fan}},  \emph {et~al.} (\bibinfo
  {collaboration} {PandaX Collaboration}),\ }\href {\doibase
  10.1103/PhysRevLett.129.161804} {\bibfield  {journal} {\bibinfo  {journal}
  {Phys. Rev. Lett.}\ }\textbf {\bibinfo {volume} {129}},\ \bibinfo {pages}
  {161804} (\bibinfo {year} {2022})}\BibitemShut {NoStop}%
\bibitem [{\citenamefont {Akerib}\ \emph {et~al.}(2021)\citenamefont {Akerib},
  \citenamefont {Al~Musalhi}, \citenamefont {Alsum}, \citenamefont
  {Amarasinghe}, \citenamefont {Ames}, \citenamefont {Anderson}, \citenamefont
  {Angelides}, \citenamefont {Ara\'ujo}, \citenamefont {Armstrong},
  \citenamefont {Arthurs} \emph {et~al.}}]{Akerib:21}%
  \BibitemOpen
  \bibfield  {author} {\bibinfo {author} {\bibfnamefont {D.~S.}\ \bibnamefont
  {Akerib}}, \bibinfo {author} {\bibfnamefont {A.~K.}\ \bibnamefont
  {Al~Musalhi}}, \bibinfo {author} {\bibfnamefont {S.~K.}\ \bibnamefont
  {Alsum}}, \bibinfo {author} {\bibfnamefont {C.~S.}\ \bibnamefont
  {Amarasinghe}}, \bibinfo {author} {\bibfnamefont {A.}~\bibnamefont {Ames}},
  \bibinfo {author} {\bibfnamefont {T.~J.}\ \bibnamefont {Anderson}}, \bibinfo
  {author} {\bibfnamefont {N.}~\bibnamefont {Angelides}}, \bibinfo {author}
  {\bibfnamefont {H.~M.}\ \bibnamefont {Ara\'ujo}}, \bibinfo {author}
  {\bibfnamefont {J.~E.}\ \bibnamefont {Armstrong}}, \bibinfo {author}
  {\bibfnamefont {M.}~\bibnamefont {Arthurs}},  \emph {et~al.},\ }\href
  {\doibase 10.1103/PhysRevD.104.092009} {\bibfield  {journal} {\bibinfo
  {journal} {Phys. Rev. D}\ }\textbf {\bibinfo {volume} {104}},\ \bibinfo
  {pages} {092009} (\bibinfo {year} {2021})}\BibitemShut {NoStop}%
\bibitem [{\citenamefont {{Carla Macolino, for the DARWIN
  collaboration}}(2020)}]{Macolino}%
  \BibitemOpen
  \bibfield  {author} {\bibinfo {author} {\bibnamefont {{Carla Macolino, for
  the DARWIN collaboration}}},\ }\href {\doibase
  10.1088/1742-6596/1468/1/012068} {\bibfield  {journal} {\bibinfo  {journal}
  {J. Phys: Conf. Ser.}\ }\textbf {\bibinfo {volume} {1468}},\ \bibinfo {pages}
  {012068} (\bibinfo {year} {2020})}\BibitemShut {NoStop}%
\bibitem [{\citenamefont {Aprile}\ \emph {et~al.}(2022)\citenamefont {Aprile},
  \citenamefont {Abe}, \citenamefont {Agostini}, \citenamefont
  {Ahmed~Maouloud}, \citenamefont {Althueser}, \citenamefont {Andrieu},
  \citenamefont {Angelino}, \citenamefont {Angevaare}, \citenamefont {Antochi},
  \citenamefont {Ant\'on~Martin} \emph {et~al.}}]{Aprile:22}%
  \BibitemOpen
  \bibfield  {author} {\bibinfo {author} {\bibfnamefont {E.}~\bibnamefont
  {Aprile}}, \bibinfo {author} {\bibfnamefont {K.}~\bibnamefont {Abe}},
  \bibinfo {author} {\bibfnamefont {F.}~\bibnamefont {Agostini}}, \bibinfo
  {author} {\bibfnamefont {S.}~\bibnamefont {Ahmed~Maouloud}}, \bibinfo
  {author} {\bibfnamefont {L.}~\bibnamefont {Althueser}}, \bibinfo {author}
  {\bibfnamefont {B.}~\bibnamefont {Andrieu}}, \bibinfo {author} {\bibfnamefont
  {E.}~\bibnamefont {Angelino}}, \bibinfo {author} {\bibfnamefont {J.~R.}\
  \bibnamefont {Angevaare}}, \bibinfo {author} {\bibfnamefont {V.~C.}\
  \bibnamefont {Antochi}}, \bibinfo {author} {\bibfnamefont {D.}~\bibnamefont
  {Ant\'on~Martin}},  \emph {et~al.} (\bibinfo {collaboration} {XENON
  Collaboration}),\ }\href {\doibase 10.1103/PhysRevLett.129.161805} {\bibfield
   {journal} {\bibinfo  {journal} {Phys. Rev. Lett.}\ }\textbf {\bibinfo
  {volume} {129}},\ \bibinfo {pages} {161805} (\bibinfo {year}
  {2022})}\BibitemShut {NoStop}%
\bibitem [{\citenamefont {Haselschwardt}\ \emph {et~al.}(2020)\citenamefont
  {Haselschwardt}, \citenamefont {Lenardo}, \citenamefont {Pirinen},\ and\
  \citenamefont {Suhonen}}]{Scott:20}%
  \BibitemOpen
  \bibfield  {author} {\bibinfo {author} {\bibfnamefont {S.}~\bibnamefont
  {Haselschwardt}}, \bibinfo {author} {\bibfnamefont {B.}~\bibnamefont
  {Lenardo}}, \bibinfo {author} {\bibfnamefont {P.}~\bibnamefont {Pirinen}}, \
  and\ \bibinfo {author} {\bibfnamefont {J.}~\bibnamefont {Suhonen}},\ }\href
  {\doibase 10.1103/PhysRevD.102.072009} {\bibfield  {journal} {\bibinfo
  {journal} {Phys. Rev. D}\ }\textbf {\bibinfo {volume} {102}},\ \bibinfo
  {pages} {072009} (\bibinfo {year} {2020})}\BibitemShut {NoStop}%
\bibitem [{\citenamefont {Frekers}\ \emph {et~al.}(2013)\citenamefont
  {Frekers}, \citenamefont {Puppe}, \citenamefont {Thies},\ and\ \citenamefont
  {Ejiri}}]{Frekers2013}%
  \BibitemOpen
  \bibfield  {author} {\bibinfo {author} {\bibfnamefont {D.}~\bibnamefont
  {Frekers}}, \bibinfo {author} {\bibfnamefont {P.}~\bibnamefont {Puppe}},
  \bibinfo {author} {\bibfnamefont {J.}~\bibnamefont {Thies}}, \ and\ \bibinfo
  {author} {\bibfnamefont {H.}~\bibnamefont {Ejiri}},\ }\href {\doibase
  https://doi.org/10.1016/j.nuclphysa.2013.08.006} {\bibfield  {journal}
  {\bibinfo  {journal} {Nucl. Phys. A}\ }\textbf {\bibinfo {volume} {916}},\
  \bibinfo {pages} {219} (\bibinfo {year} {2013})}\BibitemShut {NoStop}%
\bibitem [{\citenamefont {Bahcall}(1994)}]{Bachall:94}%
  \BibitemOpen
  \bibfield  {author} {\bibinfo {author} {\bibfnamefont {J.~N.}\ \bibnamefont
  {Bahcall}},\ }\href {\doibase 10.1103/PhysRevD.49.3923} {\bibfield  {journal}
  {\bibinfo  {journal} {Phys. Rev. D}\ }\textbf {\bibinfo {volume} {49}},\
  \bibinfo {pages} {3923} (\bibinfo {year} {1994})}\BibitemShut {NoStop}%
\bibitem [{\citenamefont {Agostini}\ \emph {et~al.}(2020)\citenamefont
  {Agostini} \emph {et~al.}}]{borexino1}%
  \BibitemOpen
  \bibfield  {author} {\bibinfo {author} {\bibfnamefont {M.}~\bibnamefont
  {Agostini}} \emph {et~al.} (\bibinfo {collaboration} {Borexino
  Collaboration}),\ }\href {\doibase 10.1038/s41586-020-2934-0} {\bibfield
  {journal} {\bibinfo  {journal} {Nature}\ }\textbf {\bibinfo {volume} {587}},\
  \bibinfo {pages} {577} (\bibinfo {year} {2020})}\BibitemShut {NoStop}%
\bibitem [{\citenamefont {Agostini}\ \emph {et~al.}(2022)\citenamefont
  {Agostini} \emph {et~al.}}]{borexino2}%
  \BibitemOpen
  \bibfield  {author} {\bibinfo {author} {\bibfnamefont {M.}~\bibnamefont
  {Agostini}} \emph {et~al.} (\bibinfo {collaboration} {Borexino
  Collaboration}),\ }\href {\doibase 10.1103/PhysRevLett.128.091803} {\bibfield
   {journal} {\bibinfo  {journal} {Phys. Rev. Lett.}\ }\textbf {\bibinfo
  {volume} {128}},\ \bibinfo {pages} {091803} (\bibinfo {year}
  {2022})}\BibitemShut {NoStop}%
\bibitem [{\citenamefont {Dror}\ \emph
  {et~al.}(2020{\natexlab{a}})\citenamefont {Dror}, \citenamefont {Elor},\ and\
  \citenamefont {McGehee}}]{Dror1}%
  \BibitemOpen
  \bibfield  {author} {\bibinfo {author} {\bibfnamefont {J.~A.}\ \bibnamefont
  {Dror}}, \bibinfo {author} {\bibfnamefont {G.}~\bibnamefont {Elor}}, \ and\
  \bibinfo {author} {\bibfnamefont {R.}~\bibnamefont {McGehee}},\ }\href
  {\doibase 10.1007/JHEP02(2020)134} {\bibfield  {journal} {\bibinfo  {journal}
  {J. High Energy Phys.}\ }\textbf {\bibinfo {volume} {2020}},\ \bibinfo
  {pages} {134} (\bibinfo {year} {2020}{\natexlab{a}})}\BibitemShut {NoStop}%
\bibitem [{\citenamefont {Dror}\ \emph
  {et~al.}(2020{\natexlab{b}})\citenamefont {Dror}, \citenamefont {Elor},\ and\
  \citenamefont {McGehee}}]{Dror2}%
  \BibitemOpen
  \bibfield  {author} {\bibinfo {author} {\bibfnamefont {J.~A.}\ \bibnamefont
  {Dror}}, \bibinfo {author} {\bibfnamefont {G.}~\bibnamefont {Elor}}, \ and\
  \bibinfo {author} {\bibfnamefont {R.}~\bibnamefont {McGehee}},\ }\href
  {\doibase 10.1103/PhysRevLett.124.181301} {\bibfield  {journal} {\bibinfo
  {journal} {Phys. Rev. Lett.}\ }\textbf {\bibinfo {volume} {124}},\ \bibinfo
  {pages} {181301} (\bibinfo {year} {2020}{\natexlab{b}})}\BibitemShut
  {NoStop}%
\bibitem [{\citenamefont {Al~Kharusi}\ \emph {et~al.}(2023)\citenamefont
  {Al~Kharusi}, \citenamefont {Anton}, \citenamefont {Badhrees}, \citenamefont
  {Barbeau}, \citenamefont {Beck}, \citenamefont {Belov}, \citenamefont
  {Bhatta}, \citenamefont {Breidenbach}, \citenamefont {Brunner}, \citenamefont
  {Cao} \emph {et~al.}}]{Soud}%
  \BibitemOpen
  \bibfield  {author} {\bibinfo {author} {\bibfnamefont {S.}~\bibnamefont
  {Al~Kharusi}}, \bibinfo {author} {\bibfnamefont {G.}~\bibnamefont {Anton}},
  \bibinfo {author} {\bibfnamefont {I.}~\bibnamefont {Badhrees}}, \bibinfo
  {author} {\bibfnamefont {P.~S.}\ \bibnamefont {Barbeau}}, \bibinfo {author}
  {\bibfnamefont {D.}~\bibnamefont {Beck}}, \bibinfo {author} {\bibfnamefont
  {V.}~\bibnamefont {Belov}}, \bibinfo {author} {\bibfnamefont
  {T.}~\bibnamefont {Bhatta}}, \bibinfo {author} {\bibfnamefont
  {M.}~\bibnamefont {Breidenbach}}, \bibinfo {author} {\bibfnamefont
  {T.}~\bibnamefont {Brunner}}, \bibinfo {author} {\bibfnamefont {G.~F.}\
  \bibnamefont {Cao}},  \emph {et~al.} (\bibinfo {collaboration} {EXO-200
  Collaboration}),\ }\href {\doibase 10.1103/PhysRevD.107.012007} {\bibfield
  {journal} {\bibinfo  {journal} {Phys. Rev. D}\ }\textbf {\bibinfo {volume}
  {107}},\ \bibinfo {pages} {012007} (\bibinfo {year} {2023})}\BibitemShut
  {NoStop}%
\bibitem [{\citenamefont {Mccutchan}(2018)}]{Mccutchan2018}%
  \BibitemOpen
  \bibfield  {author} {\bibinfo {author} {\bibfnamefont {E.~A.}\ \bibnamefont
  {Mccutchan}},\ }\href@noop {} {\bibfield  {journal} {\bibinfo  {journal}
  {Nuclear Data Sheets}\ }\textbf {\bibinfo {volume} {152}},\ \bibinfo {pages}
  {331} (\bibinfo {year} {2018})}\BibitemShut {NoStop}%
\bibitem [{\citenamefont {Wimmer}\ \emph {et~al.}(2011)\citenamefont {Wimmer},
  \citenamefont {K\"oster}, \citenamefont {Hoff}, \citenamefont {Kr\"oll},
  \citenamefont {Kr\"ucken}, \citenamefont {Lutter}, \citenamefont {Mach},
  \citenamefont {Morgan}, \citenamefont {Sarkar}, \citenamefont {Sarkar} \emph
  {et~al.}}]{Wimmer:11}%
  \BibitemOpen
  \bibfield  {author} {\bibinfo {author} {\bibfnamefont {K.}~\bibnamefont
  {Wimmer}}, \bibinfo {author} {\bibfnamefont {U.}~\bibnamefont {K\"oster}},
  \bibinfo {author} {\bibfnamefont {P.}~\bibnamefont {Hoff}}, \bibinfo {author}
  {\bibfnamefont {T.}~\bibnamefont {Kr\"oll}}, \bibinfo {author} {\bibfnamefont
  {R.}~\bibnamefont {Kr\"ucken}}, \bibinfo {author} {\bibfnamefont
  {R.}~\bibnamefont {Lutter}}, \bibinfo {author} {\bibfnamefont
  {H.}~\bibnamefont {Mach}}, \bibinfo {author} {\bibfnamefont {T.}~\bibnamefont
  {Morgan}}, \bibinfo {author} {\bibfnamefont {S.}~\bibnamefont {Sarkar}},
  \bibinfo {author} {\bibfnamefont {M.~S.}\ \bibnamefont {Sarkar}},  \emph
  {et~al.},\ }\href {\doibase 10.1103/PhysRevC.84.014329} {\bibfield  {journal}
  {\bibinfo  {journal} {Phys. Rev. C}\ }\textbf {\bibinfo {volume} {84}},\
  \bibinfo {pages} {014329} (\bibinfo {year} {2011})}\BibitemShut {NoStop}%
\bibitem [{\citenamefont {Astier}\ \emph {et~al.}(2013)\citenamefont {Astier},
  \citenamefont {Porquet}, \citenamefont {Duch\^ene}, \citenamefont {Azaiez},
  \citenamefont {Curien}, \citenamefont {Deloncle}, \citenamefont {Dorvaux},
  \citenamefont {Gall}, \citenamefont {Houry}, \citenamefont {Lucas} \emph
  {et~al.}}]{Astier:13}%
  \BibitemOpen
  \bibfield  {author} {\bibinfo {author} {\bibfnamefont {A.}~\bibnamefont
  {Astier}}, \bibinfo {author} {\bibfnamefont {M.-G.}\ \bibnamefont {Porquet}},
  \bibinfo {author} {\bibfnamefont {G.}~\bibnamefont {Duch\^ene}}, \bibinfo
  {author} {\bibfnamefont {F.}~\bibnamefont {Azaiez}}, \bibinfo {author}
  {\bibfnamefont {D.}~\bibnamefont {Curien}}, \bibinfo {author} {\bibfnamefont
  {I.}~\bibnamefont {Deloncle}}, \bibinfo {author} {\bibfnamefont
  {O.}~\bibnamefont {Dorvaux}}, \bibinfo {author} {\bibfnamefont {B.~J.~P.}\
  \bibnamefont {Gall}}, \bibinfo {author} {\bibfnamefont {M.}~\bibnamefont
  {Houry}}, \bibinfo {author} {\bibfnamefont {R.}~\bibnamefont {Lucas}},  \emph
  {et~al.},\ }\href {\doibase 10.1103/PhysRevC.87.054316} {\bibfield  {journal}
  {\bibinfo  {journal} {Phys. Rev. C}\ }\textbf {\bibinfo {volume} {87}},\
  \bibinfo {pages} {054316} (\bibinfo {year} {2013})}\BibitemShut {NoStop}%
\bibitem [{Note1()}]{Note1}%
  \BibitemOpen
  \bibinfo {note} {We believe that the authors of Ref.~\cite {Scott:20} did not
  take into account internal conversion. They quote the partial $\gamma $-ray
  transition lifetime, $\tau _\gamma = 624~\mu $s, as the total lifetime of the
  level.}\BibitemShut {Stop}%
\bibitem [{\citenamefont {Engel}\ and\ \citenamefont
  {Men{\'{e}}ndez}(2017)}]{Engel_2017}%
  \BibitemOpen
  \bibfield  {author} {\bibinfo {author} {\bibfnamefont {J.}~\bibnamefont
  {Engel}}\ and\ \bibinfo {author} {\bibfnamefont {J.}~\bibnamefont
  {Men{\'{e}}ndez}},\ }\href {\doibase 10.1088/1361-6633/aa5bc5} {\bibfield
  {journal} {\bibinfo  {journal} {Rep. Prog. Phys.}\ }\textbf {\bibinfo
  {volume} {80}},\ \bibinfo {pages} {046301} (\bibinfo {year}
  {2017})}\BibitemShut {NoStop}%
\bibitem [{\citenamefont {Ejiri}\ \emph {et~al.}(2019)\citenamefont {Ejiri},
  \citenamefont {Suhonen},\ and\ \citenamefont {Zuber}}]{Ejiri:19}%
  \BibitemOpen
  \bibfield  {author} {\bibinfo {author} {\bibfnamefont {H.}~\bibnamefont
  {Ejiri}}, \bibinfo {author} {\bibfnamefont {J.}~\bibnamefont {Suhonen}}, \
  and\ \bibinfo {author} {\bibfnamefont {K.}~\bibnamefont {Zuber}},\ }\href
  {\doibase https://doi.org/10.1016/j.physrep.2018.12.001} {\bibfield
  {journal} {\bibinfo  {journal} {Phys. Rep.}\ }\textbf {\bibinfo {volume}
  {797}},\ \bibinfo {pages} {1} (\bibinfo {year} {2019})}\BibitemShut {NoStop}%
\bibitem [{\citenamefont {Yao}\ \emph {et~al.}(2022)\citenamefont {Yao},
  \citenamefont {Meng}, \citenamefont {Niu},\ and\ \citenamefont {Ring}}]{bmf}%
  \BibitemOpen
  \bibfield  {author} {\bibinfo {author} {\bibfnamefont {J.}~\bibnamefont
  {Yao}}, \bibinfo {author} {\bibfnamefont {J.}~\bibnamefont {Meng}}, \bibinfo
  {author} {\bibfnamefont {Y.}~\bibnamefont {Niu}}, \ and\ \bibinfo {author}
  {\bibfnamefont {P.}~\bibnamefont {Ring}},\ }\href {\doibase
  https://doi.org/10.1016/j.ppnp.2022.103965} {\bibfield  {journal} {\bibinfo
  {journal} {Prog. Part. Nucl. Phys.}\ }\textbf {\bibinfo {volume} {126}},\
  \bibinfo {pages} {103965} (\bibinfo {year} {2022})}\BibitemShut {NoStop}%
\bibitem [{\citenamefont {Agostini}\ \emph {et~al.}(2023)\citenamefont
  {Agostini}, \citenamefont {Benato}, \citenamefont {Detwiler}, \citenamefont
  {Men\'endez},\ and\ \citenamefont {Vissani}}]{Agostini}%
  \BibitemOpen
  \bibfield  {author} {\bibinfo {author} {\bibfnamefont {M.}~\bibnamefont
  {Agostini}}, \bibinfo {author} {\bibfnamefont {G.}~\bibnamefont {Benato}},
  \bibinfo {author} {\bibfnamefont {J.~A.}\ \bibnamefont {Detwiler}}, \bibinfo
  {author} {\bibfnamefont {J.}~\bibnamefont {Men\'endez}}, \ and\ \bibinfo
  {author} {\bibfnamefont {F.}~\bibnamefont {Vissani}},\ }\href {\doibase
  10.1103/RevModPhys.95.025002} {\bibfield  {journal} {\bibinfo  {journal}
  {Rev. Mod. Phys.}\ }\textbf {\bibinfo {volume} {95}},\ \bibinfo {pages}
  {025002} (\bibinfo {year} {2023})}\BibitemShut {NoStop}%
\bibitem [{\citenamefont {Rebeiro}\ \emph {et~al.}(2020)\citenamefont
  {Rebeiro}, \citenamefont {Triambak}, \citenamefont {Garrett}, \citenamefont
  {Brown}, \citenamefont {Ball}, \citenamefont {Lindsay}, \citenamefont
  {Adsley}, \citenamefont {Bildstein}, \citenamefont {Burbadge}, \citenamefont
  {{Diaz Varela}} \emph {et~al.}}]{Rebeiro:20}%
  \BibitemOpen
  \bibfield  {author} {\bibinfo {author} {\bibfnamefont {B.~M.}\ \bibnamefont
  {Rebeiro}}, \bibinfo {author} {\bibfnamefont {S.}~\bibnamefont {Triambak}},
  \bibinfo {author} {\bibfnamefont {P.~E.}\ \bibnamefont {Garrett}}, \bibinfo
  {author} {\bibfnamefont {B.~A.}\ \bibnamefont {Brown}}, \bibinfo {author}
  {\bibfnamefont {G.~C.}\ \bibnamefont {Ball}}, \bibinfo {author}
  {\bibfnamefont {R.}~\bibnamefont {Lindsay}}, \bibinfo {author} {\bibfnamefont
  {P.}~\bibnamefont {Adsley}}, \bibinfo {author} {\bibfnamefont
  {V.}~\bibnamefont {Bildstein}}, \bibinfo {author} {\bibfnamefont
  {C.}~\bibnamefont {Burbadge}}, \bibinfo {author} {\bibfnamefont
  {A.}~\bibnamefont {{Diaz Varela}}},  \emph {et~al.},\ }\href {\doibase
  https://doi.org/10.1016/j.physletb.2020.135702} {\bibfield  {journal}
  {\bibinfo  {journal} {Phys. Lett. B}\ }\textbf {\bibinfo {volume} {809}},\
  \bibinfo {pages} {135702} (\bibinfo {year} {2020})}\BibitemShut {NoStop}%
\bibitem [{\citenamefont {Jokiniemi}\ \emph {et~al.}(2023)\citenamefont
  {Jokiniemi}, \citenamefont {Romeo}, \citenamefont {Soriano},\ and\
  \citenamefont {Men\'endez}}]{Lotta:23}%
  \BibitemOpen
  \bibfield  {author} {\bibinfo {author} {\bibfnamefont {L.}~\bibnamefont
  {Jokiniemi}}, \bibinfo {author} {\bibfnamefont {B.}~\bibnamefont {Romeo}},
  \bibinfo {author} {\bibfnamefont {P.}~\bibnamefont {Soriano}}, \ and\
  \bibinfo {author} {\bibfnamefont {J.}~\bibnamefont {Men\'endez}},\ }\href
  {\doibase 10.1103/PhysRevC.107.044305} {\bibfield  {journal} {\bibinfo
  {journal} {Phys. Rev. C}\ }\textbf {\bibinfo {volume} {107}},\ \bibinfo
  {pages} {044305} (\bibinfo {year} {2023})}\BibitemShut {NoStop}%
\bibitem [{\citenamefont {Cirigliano}\ \emph {et~al.}(2018)\citenamefont
  {Cirigliano}, \citenamefont {Dekens}, \citenamefont {de~Vries}, \citenamefont
  {Graesser}, \citenamefont {Mereghetti}, \citenamefont {Pastore},\ and\
  \citenamefont {van Kolck}}]{Cirigliano:18}%
  \BibitemOpen
  \bibfield  {author} {\bibinfo {author} {\bibfnamefont {V.}~\bibnamefont
  {Cirigliano}}, \bibinfo {author} {\bibfnamefont {W.}~\bibnamefont {Dekens}},
  \bibinfo {author} {\bibfnamefont {J.}~\bibnamefont {de~Vries}}, \bibinfo
  {author} {\bibfnamefont {M.~L.}\ \bibnamefont {Graesser}}, \bibinfo {author}
  {\bibfnamefont {E.}~\bibnamefont {Mereghetti}}, \bibinfo {author}
  {\bibfnamefont {S.}~\bibnamefont {Pastore}}, \ and\ \bibinfo {author}
  {\bibfnamefont {U.}~\bibnamefont {van Kolck}},\ }\href {\doibase
  10.1103/PhysRevLett.120.202001} {\bibfield  {journal} {\bibinfo  {journal}
  {Phys. Rev. Lett.}\ }\textbf {\bibinfo {volume} {120}},\ \bibinfo {pages}
  {202001} (\bibinfo {year} {2018})}\BibitemShut {NoStop}%
\bibitem [{\citenamefont {Caurier}\ \emph {et~al.}(2010)\citenamefont
  {Caurier}, \citenamefont {Nowacki}, \citenamefont {Poves},\ and\
  \citenamefont {Sieja}}]{gcn}%
  \BibitemOpen
  \bibfield  {author} {\bibinfo {author} {\bibfnamefont {E.}~\bibnamefont
  {Caurier}}, \bibinfo {author} {\bibfnamefont {F.}~\bibnamefont {Nowacki}},
  \bibinfo {author} {\bibfnamefont {A.}~\bibnamefont {Poves}}, \ and\ \bibinfo
  {author} {\bibfnamefont {K.}~\bibnamefont {Sieja}},\ }\href {\doibase
  10.1103/PhysRevC.82.064304} {\bibfield  {journal} {\bibinfo  {journal} {Phys.
  Rev. C}\ }\textbf {\bibinfo {volume} {82}},\ \bibinfo {pages} {064304}
  (\bibinfo {year} {2010})}\BibitemShut {NoStop}%
\bibitem [{\citenamefont {Men{\'{e}}ndez}(2017)}]{javier}%
  \BibitemOpen
  \bibfield  {author} {\bibinfo {author} {\bibfnamefont {J.}~\bibnamefont
  {Men{\'{e}}ndez}},\ }\href {\doibase 10.1088/1361-6471/aa9bd4} {\bibfield
  {journal} {\bibinfo  {journal} {J. Phys. G}\ }\textbf {\bibinfo {volume}
  {45}},\ \bibinfo {pages} {014003} (\bibinfo {year} {2017})}\BibitemShut
  {NoStop}%
\bibitem [{\citenamefont {Coraggio}\ \emph {et~al.}(2020)\citenamefont
  {Coraggio}, \citenamefont {Gargano}, \citenamefont {Itaco}, \citenamefont
  {Mancino},\ and\ \citenamefont {Nowacki}}]{coraggio}%
  \BibitemOpen
  \bibfield  {author} {\bibinfo {author} {\bibfnamefont {L.}~\bibnamefont
  {Coraggio}}, \bibinfo {author} {\bibfnamefont {A.}~\bibnamefont {Gargano}},
  \bibinfo {author} {\bibfnamefont {N.}~\bibnamefont {Itaco}}, \bibinfo
  {author} {\bibfnamefont {R.}~\bibnamefont {Mancino}}, \ and\ \bibinfo
  {author} {\bibfnamefont {F.}~\bibnamefont {Nowacki}},\ }\href {\doibase
  10.1103/PhysRevC.101.044315} {\bibfield  {journal} {\bibinfo  {journal}
  {Phys. Rev. C}\ }\textbf {\bibinfo {volume} {101}},\ \bibinfo {pages}
  {044315} (\bibinfo {year} {2020})}\BibitemShut {NoStop}%
\bibitem [{\citenamefont {Horoi}\ and\ \citenamefont {Brown}(2013)}]{jj55}%
  \BibitemOpen
  \bibfield  {author} {\bibinfo {author} {\bibfnamefont {M.}~\bibnamefont
  {Horoi}}\ and\ \bibinfo {author} {\bibfnamefont {B.~A.}\ \bibnamefont
  {Brown}},\ }\href {\doibase 10.1103/PhysRevLett.110.222502} {\bibfield
  {journal} {\bibinfo  {journal} {Phys. Rev. Lett.}\ }\textbf {\bibinfo
  {volume} {110}},\ \bibinfo {pages} {222502} (\bibinfo {year}
  {2013})}\BibitemShut {NoStop}%
\bibitem [{\citenamefont {Qi}\ and\ \citenamefont {Xu}(2012)}]{qx}%
  \BibitemOpen
  \bibfield  {author} {\bibinfo {author} {\bibfnamefont {C.}~\bibnamefont
  {Qi}}\ and\ \bibinfo {author} {\bibfnamefont {Z.~X.}\ \bibnamefont {Xu}},\
  }\href {\doibase 10.1103/PhysRevC.86.044323} {\bibfield  {journal} {\bibinfo
  {journal} {Phys. Rev. C}\ }\textbf {\bibinfo {volume} {86}},\ \bibinfo
  {pages} {044323} (\bibinfo {year} {2012})}\BibitemShut {NoStop}%
\bibitem [{\citenamefont {Neacsu}\ and\ \citenamefont
  {Horoi}(2015)}]{Neascu:15}%
  \BibitemOpen
  \bibfield  {author} {\bibinfo {author} {\bibfnamefont {A.}~\bibnamefont
  {Neacsu}}\ and\ \bibinfo {author} {\bibfnamefont {M.}~\bibnamefont {Horoi}},\
  }\href {\doibase 10.1103/PhysRevC.91.024309} {\bibfield  {journal} {\bibinfo
  {journal} {Phys. Rev. C}\ }\textbf {\bibinfo {volume} {91}},\ \bibinfo
  {pages} {024309} (\bibinfo {year} {2015})}\BibitemShut {NoStop}%
\bibitem [{\citenamefont {{L\"{o}ffler}}\ \emph {et~al.}(1973)\citenamefont
  {{L\"{o}ffler}}, \citenamefont {Scheerer},\ and\ \citenamefont
  {Vonach}}]{LOFFLER19731}%
  \BibitemOpen
  \bibfield  {author} {\bibinfo {author} {\bibfnamefont {M.}~\bibnamefont
  {{L\"{o}ffler}}}, \bibinfo {author} {\bibfnamefont {H.}~\bibnamefont
  {Scheerer}}, \ and\ \bibinfo {author} {\bibfnamefont {H.}~\bibnamefont
  {Vonach}},\ }\href {\doibase https://doi.org/10.1016/0029-554X(73)90090-6}
  {\bibfield  {journal} {\bibinfo  {journal} {Nucl. Instr. Meth.}\ }\textbf
  {\bibinfo {volume} {111}},\ \bibinfo {pages} {1 } (\bibinfo {year}
  {1973})}\BibitemShut {NoStop}%
\bibitem [{\citenamefont {Mukwevho}\ \emph {et~al.}(2018)\citenamefont
  {Mukwevho}, \citenamefont {Rebeiro}, \citenamefont {Mar\'{\i}n-L\'ambarri},
  \citenamefont {Triambak}, \citenamefont {Adsley}, \citenamefont {Kheswa},
  \citenamefont {Neveling}, \citenamefont {Pellegri}, \citenamefont {Pesudo},
  \citenamefont {Smit} \emph {et~al.}}]{Mukwevho2018}%
  \BibitemOpen
  \bibfield  {author} {\bibinfo {author} {\bibfnamefont {N.~J.}\ \bibnamefont
  {Mukwevho}}, \bibinfo {author} {\bibfnamefont {B.~M.}\ \bibnamefont
  {Rebeiro}}, \bibinfo {author} {\bibfnamefont {D.~J.}\ \bibnamefont
  {Mar\'{\i}n-L\'ambarri}}, \bibinfo {author} {\bibfnamefont {S.}~\bibnamefont
  {Triambak}}, \bibinfo {author} {\bibfnamefont {P.}~\bibnamefont {Adsley}},
  \bibinfo {author} {\bibfnamefont {N.~Y.}\ \bibnamefont {Kheswa}}, \bibinfo
  {author} {\bibfnamefont {R.}~\bibnamefont {Neveling}}, \bibinfo {author}
  {\bibfnamefont {L.}~\bibnamefont {Pellegri}}, \bibinfo {author}
  {\bibfnamefont {V.}~\bibnamefont {Pesudo}}, \bibinfo {author} {\bibfnamefont
  {F.~D.}\ \bibnamefont {Smit}},  \emph {et~al.},\ }\href {\doibase
  10.1103/PhysRevC.98.051302} {\bibfield  {journal} {\bibinfo  {journal} {Phys.
  Rev. C}\ }\textbf {\bibinfo {volume} {98}},\ \bibinfo {pages} {051302}
  (\bibinfo {year} {2018})}\BibitemShut {NoStop}%
\bibitem [{\citenamefont {Rebeiro}(2019)}]{BernadetteThesis}%
  \BibitemOpen
  \bibfield  {author} {\bibinfo {author} {\bibfnamefont {B.~M.}\ \bibnamefont
  {Rebeiro}},\ }\emph {\bibinfo {title} {{Nuclear structure studies in the
  $A=136$ region using transfer reactions}}},\ \href@noop {} {Ph.D. thesis},\
  \bibinfo  {school} {University of the Western Cape}, \bibinfo {address}
  {South Africa} (\bibinfo {year} {2019})\BibitemShut {NoStop}%
\bibitem [{\citenamefont {Puppe}\ \emph {et~al.}(2011)\citenamefont {Puppe}
  \emph {et~al.}}]{Puppe:11}%
  \BibitemOpen
  \bibfield  {author} {\bibinfo {author} {\bibfnamefont {P.}~\bibnamefont
  {Puppe}} \emph {et~al.},\ }\href {\doibase 10.1103/PhysRevC.84.051305}
  {\bibfield  {journal} {\bibinfo  {journal} {Phys. Rev. C}\ }\textbf {\bibinfo
  {volume} {84}},\ \bibinfo {pages} {051305} (\bibinfo {year}
  {2011})}\BibitemShut {NoStop}%
\bibitem [{\citenamefont {Rebeiro}\ \emph {et~al.}(2021)\citenamefont
  {Rebeiro}, \citenamefont {Triambak}, \citenamefont {Garrett}, \citenamefont
  {Brown}, \citenamefont {Ball}, \citenamefont {Lindsay}, \citenamefont
  {Adsley}, \citenamefont {Bildstein}, \citenamefont {Burbadge}, \citenamefont
  {Diaz-Varela} \emph {et~al.}}]{Rebeiro:21}%
  \BibitemOpen
  \bibfield  {author} {\bibinfo {author} {\bibfnamefont {B.~M.}\ \bibnamefont
  {Rebeiro}}, \bibinfo {author} {\bibfnamefont {S.}~\bibnamefont {Triambak}},
  \bibinfo {author} {\bibfnamefont {P.~E.}\ \bibnamefont {Garrett}}, \bibinfo
  {author} {\bibfnamefont {B.~A.}\ \bibnamefont {Brown}}, \bibinfo {author}
  {\bibfnamefont {G.~C.}\ \bibnamefont {Ball}}, \bibinfo {author}
  {\bibfnamefont {R.}~\bibnamefont {Lindsay}}, \bibinfo {author} {\bibfnamefont
  {P.}~\bibnamefont {Adsley}}, \bibinfo {author} {\bibfnamefont
  {V.}~\bibnamefont {Bildstein}}, \bibinfo {author} {\bibfnamefont
  {C.}~\bibnamefont {Burbadge}}, \bibinfo {author} {\bibfnamefont
  {A.}~\bibnamefont {Diaz-Varela}},  \emph {et~al.},\ }\href {\doibase
  10.1103/PhysRevC.104.034309} {\bibfield  {journal} {\bibinfo  {journal}
  {Phys. Rev. C}\ }\textbf {\bibinfo {volume} {104}},\ \bibinfo {pages}
  {034309} (\bibinfo {year} {2021})}\BibitemShut {NoStop}%
\bibitem [{\citenamefont {Glendenning}(1963)}]{Glendenning:63}%
  \BibitemOpen
  \bibfield  {author} {\bibinfo {author} {\bibfnamefont {N.~K.}\ \bibnamefont
  {Glendenning}},\ }\href {\doibase 10.1146/annurev.ns.13.120163.001203}
  {\bibfield  {journal} {\bibinfo  {journal} {Ann. Rev. Nucl. Sci.}\ }\textbf
  {\bibinfo {volume} {13}},\ \bibinfo {pages} {191} (\bibinfo {year} {1963})},\
  \Eprint
  {http://arxiv.org/abs/https://doi.org/10.1146/annurev.ns.13.120163.001203}
  {https://doi.org/10.1146/annurev.ns.13.120163.001203} \BibitemShut {NoStop}%
\bibitem [{\citenamefont {Glendenning}(1965)}]{Glendenning1965}%
  \BibitemOpen
  \bibfield  {author} {\bibinfo {author} {\bibfnamefont {N.~K.}\ \bibnamefont
  {Glendenning}},\ }\href {\doibase 10.1103/PhysRev.137.B102} {\bibfield
  {journal} {\bibinfo  {journal} {Phys. Rev.}\ }\textbf {\bibinfo {volume}
  {137}},\ \bibinfo {pages} {B102} (\bibinfo {year} {1965})}\BibitemShut
  {NoStop}%
\bibitem [{\citenamefont {An}\ and\ \citenamefont {Cai}(2006)}]{AnCai2006}%
  \BibitemOpen
  \bibfield  {author} {\bibinfo {author} {\bibfnamefont {H.}~\bibnamefont
  {An}}\ and\ \bibinfo {author} {\bibfnamefont {C.}~\bibnamefont {Cai}},\
  }\href {\doibase 10.1103/PhysRevC.73.054605} {\bibfield  {journal} {\bibinfo
  {journal} {Phys. Rev. C}\ }\textbf {\bibinfo {volume} {73}},\ \bibinfo
  {pages} {054605} (\bibinfo {year} {2006})}\BibitemShut {NoStop}%
\bibitem [{\citenamefont {Burnett}\ \emph {et~al.}(1985)\citenamefont
  {Burnett}, \citenamefont {Baxter}, \citenamefont {Hinds}, \citenamefont
  {Pribac}, \citenamefont {Smith}, \citenamefont {Spear},\ and\ \citenamefont
  {Fewell}}]{Burnett1985}%
  \BibitemOpen
  \bibfield  {author} {\bibinfo {author} {\bibfnamefont {S.~M.}\ \bibnamefont
  {Burnett}}, \bibinfo {author} {\bibfnamefont {A.~M.}\ \bibnamefont {Baxter}},
  \bibinfo {author} {\bibfnamefont {S.}~\bibnamefont {Hinds}}, \bibinfo
  {author} {\bibfnamefont {F.}~\bibnamefont {Pribac}}, \bibinfo {author}
  {\bibfnamefont {R.}~\bibnamefont {Smith}}, \bibinfo {author} {\bibfnamefont
  {R.}~\bibnamefont {Spear}}, \ and\ \bibinfo {author} {\bibfnamefont
  {M.}~\bibnamefont {Fewell}},\ }\href {\doibase
  https://doi.org/10.1016/0375-9474(85)90146-0} {\bibfield  {journal} {\bibinfo
   {journal} {Nucl. Phys. A}\ }\textbf {\bibinfo {volume} {442}},\ \bibinfo
  {pages} {289 } (\bibinfo {year} {1985})}\BibitemShut {NoStop}%
\bibitem [{Note2()}]{Note2}%
  \BibitemOpen
  \bibinfo {note} {The maximum energy of the outgoing $\alpha $'s in our
  \protect \isotope [138]Ba$(d,\alpha )$ reaction is not markedly different, at
  approximately 30~MeV.}\BibitemShut {Stop}%
\bibitem [{\citenamefont {Curry}\ \emph {et~al.}(1969)\citenamefont {Curry},
  \citenamefont {Coker},\ and\ \citenamefont {Riley}}]{Curry}%
  \BibitemOpen
  \bibfield  {author} {\bibinfo {author} {\bibfnamefont {J.~R.}\ \bibnamefont
  {Curry}}, \bibinfo {author} {\bibfnamefont {W.~R.}\ \bibnamefont {Coker}}, \
  and\ \bibinfo {author} {\bibfnamefont {P.~J.}\ \bibnamefont {Riley}},\ }\href
  {\doibase 10.1103/PhysRev.185.1416} {\bibfield  {journal} {\bibinfo
  {journal} {Phys. Rev.}\ }\textbf {\bibinfo {volume} {185}},\ \bibinfo {pages}
  {1416} (\bibinfo {year} {1969})}\BibitemShut {NoStop}%
\bibitem [{\citenamefont {{De Meijer}}\ \emph {et~al.}(1982)\citenamefont {{De
  Meijer}}, \citenamefont {Put}, \citenamefont {Akkerman}, \citenamefont
  {Vermeulen},\ and\ \citenamefont {Bingham}}]{Rob1}%
  \BibitemOpen
  \bibfield  {author} {\bibinfo {author} {\bibfnamefont {R.~J.}\ \bibnamefont
  {{De Meijer}}}, \bibinfo {author} {\bibfnamefont {L.~W.}\ \bibnamefont
  {Put}}, \bibinfo {author} {\bibfnamefont {J.~J.}\ \bibnamefont {Akkerman}},
  \bibinfo {author} {\bibfnamefont {J.~C.}\ \bibnamefont {Vermeulen}}, \ and\
  \bibinfo {author} {\bibfnamefont {C.~R.}\ \bibnamefont {Bingham}},\ }\href
  {\doibase https://doi.org/10.1016/0375-9474(82)90111-7} {\bibfield  {journal}
  {\bibinfo  {journal} {Nucl. Phys. A}\ }\textbf {\bibinfo {volume} {386}},\
  \bibinfo {pages} {200} (\bibinfo {year} {1982})}\BibitemShut {NoStop}%
\bibitem [{\citenamefont {Daehnick}\ and\ \citenamefont
  {Park}(1969)}]{DaehnickPark1969}%
  \BibitemOpen
  \bibfield  {author} {\bibinfo {author} {\bibfnamefont {W.~W.}\ \bibnamefont
  {Daehnick}}\ and\ \bibinfo {author} {\bibfnamefont {Y.~S.}\ \bibnamefont
  {Park}},\ }\href {\doibase 10.1103/PhysRev.180.1062} {\bibfield  {journal}
  {\bibinfo  {journal} {Phys. Rev.}\ }\textbf {\bibinfo {volume} {180}},\
  \bibinfo {pages} {1062} (\bibinfo {year} {1969})}\BibitemShut {NoStop}%
\bibitem [{\citenamefont {Charlton}(1973)}]{Charlton}%
  \BibitemOpen
  \bibfield  {author} {\bibinfo {author} {\bibfnamefont {L.~A.}\ \bibnamefont
  {Charlton}},\ }\href {\doibase 10.1103/PhysRevC.8.146} {\bibfield  {journal}
  {\bibinfo  {journal} {Phys. Rev. C}\ }\textbf {\bibinfo {volume} {8}},\
  \bibinfo {pages} {146} (\bibinfo {year} {1973})}\BibitemShut {NoStop}%
\bibitem [{\citenamefont {DelVecchio}\ and\ \citenamefont
  {Daehnick}(1972)}]{DelVecchio1972}%
  \BibitemOpen
  \bibfield  {author} {\bibinfo {author} {\bibfnamefont {R.~M.}\ \bibnamefont
  {DelVecchio}}\ and\ \bibinfo {author} {\bibfnamefont {W.~W.}\ \bibnamefont
  {Daehnick}},\ }\href {\doibase 10.1103/PhysRevC.6.2095} {\bibfield  {journal}
  {\bibinfo  {journal} {Phys. Rev. C}\ }\textbf {\bibinfo {volume} {6}},\
  \bibinfo {pages} {2095} (\bibinfo {year} {1972})}\BibitemShut {NoStop}%
\bibitem [{\citenamefont {Brown}\ \emph {et~al.}(2005)\citenamefont {Brown},
  \citenamefont {Stone}, \citenamefont {Stone}, \citenamefont {Towner},\ and\
  \citenamefont {Hjorth-Jensen}}]{sn100}%
  \BibitemOpen
  \bibfield  {author} {\bibinfo {author} {\bibfnamefont {B.~A.}\ \bibnamefont
  {Brown}}, \bibinfo {author} {\bibfnamefont {N.~J.}\ \bibnamefont {Stone}},
  \bibinfo {author} {\bibfnamefont {J.~R.}\ \bibnamefont {Stone}}, \bibinfo
  {author} {\bibfnamefont {I.~S.}\ \bibnamefont {Towner}}, \ and\ \bibinfo
  {author} {\bibfnamefont {M.}~\bibnamefont {Hjorth-Jensen}},\ }\href {\doibase
  10.1103/PhysRevC.71.044317} {\bibfield  {journal} {\bibinfo  {journal} {Phys.
  Rev. C}\ }\textbf {\bibinfo {volume} {71}},\ \bibinfo {pages} {044317}
  (\bibinfo {year} {2005})}\BibitemShut {NoStop}%
\bibitem [{\citenamefont {Brown}\ \emph {et~al.}(2014)\citenamefont {Brown},
  \citenamefont {Horoi},\ and\ \citenamefont {Sen'kov}}]{Brown:14}%
  \BibitemOpen
  \bibfield  {author} {\bibinfo {author} {\bibfnamefont {B.~A.}\ \bibnamefont
  {Brown}}, \bibinfo {author} {\bibfnamefont {M.}~\bibnamefont {Horoi}}, \ and\
  \bibinfo {author} {\bibfnamefont {R.~A.}\ \bibnamefont {Sen'kov}},\ }\href
  {\doibase 10.1103/PhysRevLett.113.262501} {\bibfield  {journal} {\bibinfo
  {journal} {Phys. Rev. Lett.}\ }\textbf {\bibinfo {volume} {113}},\ \bibinfo
  {pages} {262501} (\bibinfo {year} {2014})}\BibitemShut {NoStop}%
\bibitem [{\citenamefont {Dabbousi}\ \emph {et~al.}(1971)\citenamefont
  {Dabbousi}, \citenamefont {Prior},\ and\ \citenamefont {Shugart}}]{Dabb:71}%
  \BibitemOpen
  \bibfield  {author} {\bibinfo {author} {\bibfnamefont {O.~B.}\ \bibnamefont
  {Dabbousi}}, \bibinfo {author} {\bibfnamefont {M.~H.}\ \bibnamefont {Prior}},
  \ and\ \bibinfo {author} {\bibfnamefont {H.~A.}\ \bibnamefont {Shugart}},\
  }\href {\doibase 10.1103/PhysRevC.3.1326} {\bibfield  {journal} {\bibinfo
  {journal} {Phys. Rev. C}\ }\textbf {\bibinfo {volume} {3}},\ \bibinfo {pages}
  {1326} (\bibinfo {year} {1971})}\BibitemShut {NoStop}%
\bibitem [{\citenamefont {Thibault}\ \emph {et~al.}(1981)\citenamefont
  {Thibault}, \citenamefont {Touchard}, \citenamefont {B\"uttgenbach},
  \citenamefont {Klapisch}, \citenamefont {{De Saint Simon}}, \citenamefont
  {Duong}, \citenamefont {Jacquinot}, \citenamefont {Juncar}, \citenamefont
  {Liberman}, \citenamefont {Pillet} \emph {et~al.}}]{Thibault:81}%
  \BibitemOpen
  \bibfield  {author} {\bibinfo {author} {\bibfnamefont {C.}~\bibnamefont
  {Thibault}}, \bibinfo {author} {\bibfnamefont {F.}~\bibnamefont {Touchard}},
  \bibinfo {author} {\bibfnamefont {S.}~\bibnamefont {B\"uttgenbach}}, \bibinfo
  {author} {\bibfnamefont {R.}~\bibnamefont {Klapisch}}, \bibinfo {author}
  {\bibfnamefont {M.}~\bibnamefont {{De Saint Simon}}}, \bibinfo {author}
  {\bibfnamefont {H.}~\bibnamefont {Duong}}, \bibinfo {author} {\bibfnamefont
  {P.}~\bibnamefont {Jacquinot}}, \bibinfo {author} {\bibfnamefont
  {P.}~\bibnamefont {Juncar}}, \bibinfo {author} {\bibfnamefont
  {S.}~\bibnamefont {Liberman}}, \bibinfo {author} {\bibfnamefont
  {P.}~\bibnamefont {Pillet}},  \emph {et~al.},\ }\href {\doibase
  https://doi.org/10.1016/0375-9474(81)90274-8} {\bibfield  {journal} {\bibinfo
   {journal} {Nucl. Phys. A}\ }\textbf {\bibinfo {volume} {367}},\ \bibinfo
  {pages} {1} (\bibinfo {year} {1981})}\BibitemShut {NoStop}%
\bibitem [{\citenamefont {Kamil}\ \emph {et~al.}(2022)\citenamefont {Kamil},
  \citenamefont {Triambak}, \citenamefont {Ball}, \citenamefont {Bildstein},
  \citenamefont {Varela}, \citenamefont {Faestermann}, \citenamefont {Garrett},
  \citenamefont {Moradi}, \citenamefont {Hertenberger}, \citenamefont
  {Kheswa},\ and\ \citenamefont {ohers}}]{Kamil:22}%
  \BibitemOpen
  \bibfield  {author} {\bibinfo {author} {\bibfnamefont {M.}~\bibnamefont
  {Kamil}}, \bibinfo {author} {\bibfnamefont {S.}~\bibnamefont {Triambak}},
  \bibinfo {author} {\bibfnamefont {G.~C.}\ \bibnamefont {Ball}}, \bibinfo
  {author} {\bibfnamefont {V.}~\bibnamefont {Bildstein}}, \bibinfo {author}
  {\bibfnamefont {A.~D.}\ \bibnamefont {Varela}}, \bibinfo {author}
  {\bibfnamefont {T.}~\bibnamefont {Faestermann}}, \bibinfo {author}
  {\bibfnamefont {P.~E.}\ \bibnamefont {Garrett}}, \bibinfo {author}
  {\bibfnamefont {F.~G.}\ \bibnamefont {Moradi}}, \bibinfo {author}
  {\bibfnamefont {R.}~\bibnamefont {Hertenberger}}, \bibinfo {author}
  {\bibfnamefont {N.~Y.}\ \bibnamefont {Kheswa}}, \ and\ \bibinfo {author}
  {\bibnamefont {ohers}},\ }\href {\doibase 10.1103/PhysRevC.105.055805}
  {\bibfield  {journal} {\bibinfo  {journal} {Phys. Rev. C}\ }\textbf {\bibinfo
  {volume} {105}},\ \bibinfo {pages} {055805} (\bibinfo {year}
  {2022})}\BibitemShut {NoStop}%
\bibitem [{\citenamefont {Gimeno}\ \emph {et~al.}(2023)\citenamefont {Gimeno},
  \citenamefont {Jokiniemi}, \citenamefont {Kotila}, \citenamefont {Ramalho},\
  and\ \citenamefont {Suhonen}}]{Gimeno:23}%
  \BibitemOpen
  \bibfield  {author} {\bibinfo {author} {\bibfnamefont {P.}~\bibnamefont
  {Gimeno}}, \bibinfo {author} {\bibfnamefont {L.}~\bibnamefont {Jokiniemi}},
  \bibinfo {author} {\bibfnamefont {J.}~\bibnamefont {Kotila}}, \bibinfo
  {author} {\bibfnamefont {M.}~\bibnamefont {Ramalho}}, \ and\ \bibinfo
  {author} {\bibfnamefont {J.}~\bibnamefont {Suhonen}},\ }\href {\doibase
  10.3390/universe9060270} {\bibfield  {journal} {\bibinfo  {journal}
  {Universe}\ }\textbf {\bibinfo {volume} {9}} (\bibinfo {year} {2023}),\
  10.3390/universe9060270}\BibitemShut {NoStop}%
\bibitem [{Note3()}]{Note3}%
  \BibitemOpen
  \bibinfo {note} {A careful analysis of our $\alpha $ spectrum rules out any
  possible intermediate $4^+$ state below 74~keV, which would enable faster
  $M1$ transitions and lead to a substantially shorter lifetime for the $3_1^+$
  level.}\BibitemShut {Stop}%
\bibitem [{ic()}]{ic}%
  \BibitemOpen
  \href@noop {} {}\bibinfo {howpublished}
  {\url{https://bricc.anu.edu.au/}}\BibitemShut {NoStop}%
\bibitem [{\citenamefont {Haselschwardt}\ \emph {et~al.}(2023)\citenamefont
  {Haselschwardt}, \citenamefont {Lenardo}, \citenamefont {Daniels},
  \citenamefont {Finch}, \citenamefont {Friesen}, \citenamefont {Howell},
  \citenamefont {Malone}, \citenamefont {Mancil},\ and\ \citenamefont
  {Tornow}}]{tunl}%
  \BibitemOpen
  \bibfield  {author} {\bibinfo {author} {\bibfnamefont {S.~J.}\ \bibnamefont
  {Haselschwardt}}, \bibinfo {author} {\bibfnamefont {B.~G.}\ \bibnamefont
  {Lenardo}}, \bibinfo {author} {\bibfnamefont {T.}~\bibnamefont {Daniels}},
  \bibinfo {author} {\bibfnamefont {S.~W.}\ \bibnamefont {Finch}}, \bibinfo
  {author} {\bibfnamefont {F.~Q.}\ \bibnamefont {Friesen}}, \bibinfo {author}
  {\bibfnamefont {C.~R.}\ \bibnamefont {Howell}}, \bibinfo {author}
  {\bibfnamefont {C.~R.}\ \bibnamefont {Malone}}, \bibinfo {author}
  {\bibfnamefont {E.}~\bibnamefont {Mancil}}, \ and\ \bibinfo {author}
  {\bibfnamefont {W.}~\bibnamefont {Tornow}},\ }\href@noop {} {\enquote
  {\bibinfo {title} {Observation of low-lying isomeric states in {$^{136}$Cs}:
  a new avenue for dark matter and solar neutrino detection in xenon
  detectors},}\ } (\bibinfo {year} {2023}),\ \Eprint
  {http://arxiv.org/abs/2301.11893} {arXiv:2301.11893 [nucl-ex]} \BibitemShut
  {NoStop}%
\end{thebibliography}%
% 
%\newpage

\end{document}